\documentclass{aa}  

\usepackage{graphicx}
\usepackage{txfonts}
\usepackage{physics}
\usepackage{booktabs}
\usepackage{multirow}
\usepackage{url}
\usepackage{xcolor}
\defcitealias{ShiKremerGrudic_2023MNRAS.518.3606S}{Paper~I}
\begin{document}

   \title{Feedback-regulated Seed Black Hole Growth in Star-Forming Molecular Clouds and Galactic Nuclei}


   \author{Yanlong Shi
          \inst{1,2} \thanks{E-mail: yanlong.astro@outlook.com}
          \and
          Kyle Kremer \inst{1}
          \and
          Philip F. Hopkins \inst{1}
          }

   \institute{TAPIR, MC 350-17, California Institute of Technology, Pasadena, CA 91125, USA
   		\and
   			Canadian Institute for Theoretical Astrophysics, University of Toronto, Toronto, ON M5S 3H8, Canada
             }

   \date{Received ; accepted }


  \abstract
   {The detection of supermassive black holes (SMBHs) in high-redshift luminous quasars may require a phase of rapid accretion, and as a precondition, substantial gas influx toward seed black holes (BHs) from kilo-parsec or parsec scales. Our previous research demonstrated the plausibility of such gas supply for BH seeds within star-forming giant molecular clouds (GMCs) with high surface density ($\sim 10^4\,{\rm {\rm M_\odot}\, pc}^{-2}$), facilitating ``hyper-Eddington'' accretion via efficient feeding by dense clumps which are driven by turbulence and stellar feedback.}
   {This article investigates the impacts of feedback from accreting BHs on this process, including radiation, mechanical jets, and highly relativistic cosmic rays. }
   {We run a suite of numerical simulations to explore diverse parameter spaces of BH feedback, including the sub-grid accretion model, feedback energy efficiency, mass loading factor, and initial metallicity.}
   {Utilizing radiative feedback models inferred from the slim disk, we find that hyper-Eddington accretion is still achievable, yielding BH bolometric luminosities as high as $10^{41}$ -- $10^{44}\,\rm erg/s$, depending on the GMC properties and specific feedback model assumed. We find the maximum possible mass growth of seed BHs ($\Delta M_{\rm BH}^{\rm max}$) is regulated by the momentum deposition rate from BH feedback, $\dot{p}_{\rm feedback}/(\dot{M}_{\rm BH} c)$, which leads to an analytic scaling that agrees well with simulations. This scenario predicts the rapid formation of $\sim 10^4\,\rm M_\odot$ intermediate-massive BHs (IMBHs) from stellar-mass BHs within $\sim \rm Myr$. Furthermore, we examine the impacts of sub-grid accretion models and how BH feedback may influence star formation within these cloud complexes.}
   {}


   \maketitle
%

\nolinenumbers

\section{Introduction}
\label{sec:introduction}

Formation of supermassive black holes (SMBHs), especially those weighing $\sim 10^9\,{\rm M_\odot}$ at very high redshift like $z \gtrsim 7$ \citep[e.g., recent observations by][]{BanadosVenemansMazzucchelli_2018Natur.553..473B,YangWangFan_2020ApJ...897L..14Y,WangYangFan_2021ApJ...907L...1W}, has long been an intriguing astrophysical problem \citep[e.g., recent reviews by][]{InayoshiVisbalHaiman_2020ARA&A..58...27I,VolonteriHabouzitColpi_2021NatRP...3..732V},. Observations have shown that these SMBHs grow from lighter ``seed'' BHs \citep[]{YuTremaine_2002MNRAS.335..965Y}. The seeds, typically in the mass range of intermediate-mass black holes (IMBHs; $\sim 100$ -- $10^6\,{\rm M_\odot}$), are proposed to form in multiple astrophysical scenarios like the direct collapse of the pristine gas \citep[e.g.,][]{BrommLoeb_2003ApJ...596...34B,LatifWhalenKhochfar_2022Natur.607...48L}, Population III star remnants \citep[e.g.][]{MadauRees_2001ApJ...551L..27M,RyuTanakaPerna_2016MNRAS.460.4122R}, stellar-mass BHs that undergo hyper-Eddington accretion \citep[e.g.,][]{LupiHaardtDotti_2016MNRAS.456.2993L,PezzulliValianteSchneider_2016MNRAS.458.3047P,ReganDownesVolonteri_2019MNRAS.486.3892R,ShiKremerGrudic_2023MNRAS.518.3606S,LupiQuadriVolonteri_2024A&A...686A.256L}, and runaway stellar mergers in dense star clusters \citep[e.g.,][]{PortegiesZwartBaumgardtHut_2004Natur.428..724P,KremerSperaBecker_2020ApJ...903...45K,ShiGrudicHopkins_2021MNRAS.505.2753S}. If we use a reference radiative efficiency of 0.1, the $e$-folding time scale at the Eddington accretion rate is $\sim 45\,{\rm Myr}$; on the other hand, due to the limited time allowed for SMBH formation (especially for $z\gtrsim 7$ quasars that formed only $\lesssim 10^9\,{\rm yr}$ after the Big Bang), a sustained phase of fast accretion is inevitable, possibly at super- or hyper-Eddington ($\gtrsim 500\,\dot{M}_{\rm Edd}$) accretion rates \citep{InayoshiVisbalHaiman_2020ARA&A..58...27I}.\footnote{
	Here $\dot{M}_{\rm Edd} \equiv M_{\rm BH}/t_{\rm Sal}$, where $t_{\rm Sal}=0.1 \sigma_{\rm T} c/(4\pi G\,m_{\rm p})\approx 45\,{\rm Myr}$.
}

A number of theoretical works on small-scale BH accretion physics have shown that super-Eddington accretion is achievable and sustainable in principle within the BH accretion disk \citep[]{Begelman_1979MNRAS.187..237B,QuataertGruzinov_2000ApJ...539..809Q,BlandfordBegelman_2004MNRAS.349...68B,InayoshiHaimanOstriker_2016MNRAS.459.3738I}. This is also supported by simulations with RHD \citep[]{OhsugaMoriNakamoto_2005ApJ...628..368O}, RMHD \citep[]{JiangStoneDavis_2014ApJ...796..106J,JiangStoneDavis_2019ApJ...880...67J} and GRMHD \citep[]{SadowskiNarayanTchekhovskoy_2015MNRAS.447...49S}. However, these simulations typically embed the BH inside a gas reservoir with a sufficiently high mass supply rate from large radii ($\gtrsim 10^3 \dot{M}_{\rm Edd}$). This level of gas feeding is \textit{not} necessarily possible in more realistic star-forming (much ``larger-scale'' compared to BH disks) astrophysical environments, since the gas medium may fragment due to self-gravity and result in star formation; and feedback from newly-formed stars, including radiative pressure, photoionization/heating, and winds, will develop turbulence or bulk motion in the medium \citep[e.g., see star formation simulations][]{GrudicGuszejnovHopkins_2018MNRAS.481..688G,GrudicGuszejnovHopkins_2021MNRAS.506.2199G}. Strong stellar feedback near an accreting BH may deplete the {available} fuel and challenge the {feasibility} of super-Eddington accretion \citep[]{DuboisVolonteriSilk_2015MNRAS.452.1502D,HabouzitVolonteriDubois_2017MNRAS.468.3935H,BowerSchayeFrenk_2017MNRAS.465...32B}. Light seeds, like IMBHs or stellar-mass BHs, are especially prone to {the accretion challenges inherent to} these environments since regimes influenced/dominated by their gravity \citep[i.e., the Bondi-Hoyle radius,][]{HoyleLyttleton_1939PCPS...35..405H,Bondi_1952MNRAS.112..195B} are small compared with galactic scales.

To reconcile the problem, ``larger-scale'' (kpc or pc scale) simulations are not only required to model star formation and feedback in sufficient details, but also to resolve the some gas dynamics \textit{below} the gravitational capture scale (Bondi-Hoyle radius) of at least some rapid-accreting BHs. In \citet{ShiKremerGrudic_2023MNRAS.518.3606S} (\citetalias{ShiKremerGrudic_2023MNRAS.518.3606S} hereafter), we made some first steps towards addressing this problem, using simulations of star formation and feedback in giant molecular clouds (GMCs) similar to that in \cite{GrudicGuszejnovHopkins_2018MNRAS.481..688G}, and studying BH seed growth in the absence of BH feedback. Although the stellar feedback makes the GMC uneven and turbulent, it also generates dense clumps and shocks \citep[]{Klessen_2000ApJ...535..869K}, which may have low internal velocity dispersion \citep[]{MacLowKlessen_2004RvMP...76..125M,McKeeOstriker_2007ARA&A..45..565M} -- these factors make the simulation possible to resolve the gravitational capture of gas for some ``lucky'' BHs encountering dense gas clumps with small relative velocity, since the Bondi-Hyle radii are large under these circumstances. Similar to the star formation efficiency explored in the literature \citep[]{GrudicGuszejnovHopkins_2018MNRAS.481..688G,KimKimOstriker_2018ApJ...859...68K,HeRicottiGeen_2019MNRAS.489.1880H,FukushimaYajimaSugimura_2020MNRAS.497.3830F,GrudicGuszejnovHopkins_2021MNRAS.506.2199G,KimOstrikerFilippova_2021ApJ...911..128K,ChevanceKrumholzMcLeod_2023ASPC..534....1C}, BH accretion is also regulated by the surface density ($\Sigma_0\equiv M_{\rm cl}/\pi R_{\rm cl}^2$) of the cloud: higher $\Sigma_0$ means higher self-gravity ($\sim G M_{\rm cl}/R_{\rm cl}^2 \sim \pi G \Sigma_0$) which keeps the cloud bound rather than disrupted by stellar feedback. \citetalias{ShiKremerGrudic_2023MNRAS.518.3606S} found that for clouds with high surface density ($\Sigma_0 \approx 10^4\,{\rm {\rm M_\odot} \,pc^{-2}}$), the gas supply reaching the BH's accretion disk could be {sufficient} to support hyper-Eddington accretion with $f_{\rm Edd} \equiv \dot{M}_{\rm BH}/\dot{M}_{\rm Edd} \sim 1000$ for at least a small but non-negligible amount of seeds.

More realistically, BH feedback mechanisms, like electromagnetic radiation \citep[]{Fabian_2012ARA&A..50..455F,HopkinsTorreyFaucher-Giguere_2016MNRAS.458..816H}, winds/jets \citep[]{SilkRees_1998A&A...331L...1S,MurrayQuataertThompson_2005ApJ...618..569M,DiMatteoSpringelHernquist_2005Natur.433..604D,OstrikerChoiCiotti_2010ApJ...722..642O,SadowskiLasotaAbramowicz_2016MNRAS.456.3915S,TorreyHopkinsFaucher-Giguere_2020MNRAS.497.5292T,HuInayoshiHaiman_2022ApJ...934..132H,MassonneauVolonteriDubois_2023A&A...670A.180M}, and cosmic rays \citep[CRs;][]{SijackiPfrommerSpringel_2008MNRAS.387.1403S,GuoMathews_2012ApJ...756..181G,Zweibel_2017PhPl...24e5402Z,SuHopkinsBryan_2021MNRAS.507..175S,IshibashiFabian_2023MNRAS.519.1931I} will deposit energy and mechanical momentum back to the gas reservoir. The energy and momentum heat and deplete the gas fuel,  or even prevent its infall to the BH \citep[]{DuboisPichonDevriendt_2013MNRAS.428.2885D}, {potentially} challenging the hyper-Eddington accretion scenario. Still, close to the BH ($\lesssim 1000\,r_{\rm Sch}$), at {sufficiently high accretion rates and opacities}, photons are effectively trapped with the inflow, which means the radiated energy \citep[and other references above]{WataraiFukueTakeuchi_2000PASJ...52..133W,MadauHaardtDotti_2014ApJ...784L..38M} will only grow sub-linearly with the accretion rate (measured as $\dot{M}_{\rm BH}/\dot{M}_{\rm Edd}$). This is also true for the mass outflow powered by radiation \citep[]{HuInayoshiHaiman_2022ApJ...934..132H}. As a result, the feedback strength at high accretion rates is suppressed and hyper-Eddington accretion could remain possible.

Though BH feedback is not included in the simulations of \citetalias{ShiKremerGrudic_2023MNRAS.518.3606S}, there is an estimate in terms of radiative feedback based on the density criteria for hyper-Eddington accretion predicted by \cite{InayoshiHaimanOstriker_2016MNRAS.459.3738I}: $n_{\infty} \gtrsim 10^{5}(M_{\rm BH}/10^4\,{\rm M_\odot})^{-1}(T_{\infty}/10^4\,{\rm K})^{3/2}\,{\rm cm^{-3}}$, where $n_{\infty}$ and $T_\infty$ are the density and temperature of gas ``near'' the BH (at sub-pc scales). For BHs in simulations that grow rapidly, especially those inside GMCs with $\Sigma_0\approx 10^4\,{\rm M_\odot \,pc^{-2}}$, the ambient gas density is above this critical value, which {suggests} hyper-Eddington accretion may be achievable even if there is radiative feedback. Still, there is a necessity to include BH radiative feedback in simulations like \citetalias{ShiKremerGrudic_2023MNRAS.518.3606S} for completeness since BH feedback may still affect GMC behaviors at larger scales. The influence of other forms of BH feedback, {namely, mechanical feedback and cosmic rays} on accretion is also to be determined.

In this article, we extend the discussion of \citetalias{ShiKremerGrudic_2023MNRAS.518.3606S} and explore the impact of multiple BH feedback mechanisms, including radiation, mechanical outflow, and CRs. In particular, we are interested in how the BH feedback mechanisms determine the maximum mass a BH can reach from accretion, and how they scale in different GMCs. To reach this goal, we run a suite of numerical simulations by adopting the basic setups and physics involved in \citetalias{ShiKremerGrudic_2023MNRAS.518.3606S} where the authors showed hyper-Eddington accretion was possible \textit{without} BH feedback, but with a new sub-grid BH accretion and feedback model, as well as related physics. 

The article is organized as follows. In \S~\ref{sec:method}, we introduce the background and implementations of BH accretion and feedback, as well as the simulation initial conditions. In \S~\ref{sec:results} , we present the result with the fiducial feedback models and explore the effects of different accretion/feedback physics. In \S~\ref{sec:discussions}, we discuss the result and important caveats of our simulations. Finally, we conclude in \S~\ref{sec:conclusions}.

\begin{figure}
	\centering
	\includegraphics[width=\linewidth]{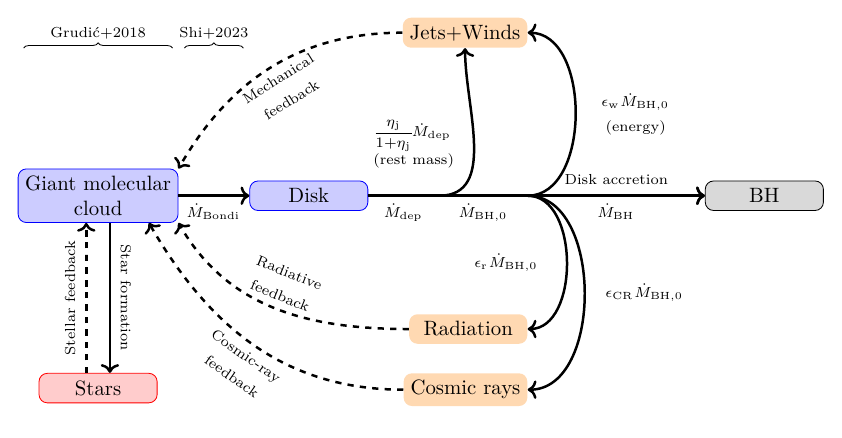}
	\vspace{-10 pt}
	\caption{Mass and energy flow of BH accretion and feedback. This study extends the framework of star formation in \citet{GrudicGuszejnovHopkins_2018MNRAS.481..688G} and BH accretion without feedback in \citetalias{ShiKremerGrudic_2023MNRAS.518.3606S}. To implement BH feedback, we use a model of mass transfer from the disk to the BH. With only a fraction of the accretion flow going to the BH, other portions are deposited back to the GMC in the form of radiation, jets/winds, and cosmic rays. \textcolor{black}{We note that since the feedback energy efficiencies are typically small ($\lesssim 0.1$), $\dot M_{\rm BH} \doteq \dot M_{\rm BH,0}$. }}
	\label{fig:flowchart}
\end{figure}

\section{Background and Method}
\label{sec:method}

\begin{table*}
	\centering
	\caption{Initial conditions involved in this study. We chose some gas complexes in the ``high surface-density'' ($\Sigma_0=13000\,{\rm M_\odot\,pc^{-2}}$) group of \citetalias{ShiKremerGrudic_2023MNRAS.518.3606S}, with the initial radius ($R_{\rm cl}$) of $5\,{\rm pc}$ or $50\,{\rm pc}$. Each kind of gas complex is simulated at high or low resolution. Starting from the third column, we show the initial free-fall time ($t_{\rm ff}$), gas mass resolution ($m_{\rm gas}$),  softening radius for stars ($r_{\rm soft}^{\rm star}$) and BHs ($r_{\rm soft}^{\rm BH}$), characteristic jet energy efficiency ($\epsilon_{\rm w,c}$) and velocity ($v_{\rm w,c}$) as described in \S\ref{sec:method}, number of BH seeds ($N_{\rm BH}$), and initial BH mass ($M_{\rm BH}^{\rm ini}$) for different kinds of simulations.
	}
	\begin{tabular}{llllllllllll}
		\toprule
		$M_{\rm cl}\,[{\rm M_\odot}]$ & $R_{\rm cl}~[{\rm pc}]$ &$t_{\rm ff}\,[{\rm Myr}]$  & $m_{\rm gas}\,[{\rm M_\odot}]$ & $r^{\rm star}_{\rm soft}\,[{\rm pc}]$ & $r^{\rm BH}_{\rm soft}\,[{\rm pc}]$  & $\epsilon_{\rm w, c}$ & $v_{\rm w,c}\,[{\rm km/s}]$ & $N_{\rm BH}$& $M_{\rm BH}^{\rm ini}\,[\rm M_\odot]$ & Notes \\
		\midrule
		$10^6$ &5 & 0.19 &3.8 & 0.41 & 0.31 & $10^{-6}$ & 420 & 234 & 10 -- $10^3$ & Low res.\\
		$10^6$ &5 & 0.19 &0.48 & 0.21 & 0.21 & $10^{-6}$ & 420 & 234 &10 -- $10^3$ & High res.\\
		\midrule
		$10^8$ &50 & 0.59 &380 & 1.9 & 0.31 & $10^{-3}$ & 13000 & 2000 & $10^2$ -- $10^4$ & Low res.\\
		$10^8$ &50 & 0.59 &48 & 0.96 & 0.31 & $10^{-3}$ & 13000 & 2000 & $10^2$ -- $10^4$ & High res.\\
		\bottomrule
	\end{tabular}
	\label{tab:gmc_setups}
\end{table*}

This study follows the basic framework of \citetalias{ShiKremerGrudic_2023MNRAS.518.3606S}: each simulation is initialized with a number of seed BHs inside a pre-collapse pre-star-formation parsec-scale GMC, and then evolves self-consistently with MHD+gravity and additional physics including star formation/feedback, BH accretion/feedback, etc. 

\subsection{Overview}

Our simulation follows the same numerical framework for star formation and feedback as that in previous star-formation simulations \citep[]{GrudicGuszejnovHopkins_2018MNRAS.481..688G,GrudicKruijssenFaucher-Giguere_2021MNRAS.506.3239G} and the previous study of this series \citepalias{ShiKremerGrudic_2023MNRAS.518.3606S}. As a short summary, we use the meshless, Lagrangian, Godunov MHD code GIZMO \citep[]{Hopkins_2015MNRAS.450...53H,Hopkins_2016MNRAS.462..576H,HopkinsRaives_2016MNRAS.455...51H} in its Meshless Finite Mass (MFM) mode, including the physics of self-gravity, radiative cooling, star formation, and feedback. In particular, star formation and feedback are based on the FIRE-3 implementation of the Feedback In Realistic Environments\footnote{\url{http://fire.northwestern.edu/}} (FIRE) framework \citep[]{HopkinsWetzelKeres_2018MNRAS.480..800H,HopkinsWetzelWheeler_2023MNRAS.519.3154H}. Each ``star'' particle in the simulation represents an IMF-averaged ensemble of stars, which evolves and deposits mass outflow, metals, and radiation back to the ISM as its feedback \citep[]{HopkinsWetzelKeres_2018MNRAS.477.1578H}. This treatment has been shown to reasonably reproduce many of the properties of star formation GMCs, like the star formation efficiency, star cluster dynamics, and cluster mass distribution \citep[]{GrudicGuszejnovHopkins_2018MNRAS.481..688G,GrudicKruijssenFaucher-Giguere_2021MNRAS.506.3239G}.

The simulation also follows the same scheme for BH accretion from gravitational capture as that in \citetalias{ShiKremerGrudic_2023MNRAS.518.3606S}, based on that stated in \citet{GrudicGuszejnovHopkins_2021MNRAS.506.2199G}: gas cells are captured by a BH if (i) they are within a preset ``sink radius'' near the BH, (ii) they are gravitationally bound to the BH, and (iii) their individual Keplerian orbit is within the sink radius \citep[]{BateBonnellPrice_1995MNRAS.277..362B}. Moreover, we choose the softening radius of BHs to be the same as its sink radius and set the value to make sure that the Bondi-Hoyle accretion radius is resolved (i.e., greater than the sink radius) for fast-accreting BHs, which is extensively discussed in \citetalias{ShiKremerGrudic_2023MNRAS.518.3606S}.

For this study, we extend the BH accretion scheme with a sub-grid model of BH accretion to account for BH feedback. To help explain this, we present a schematic view of the simulation in Fig.~\ref{fig:flowchart}. Accretion onto a BH can be decomposed into two successive phases: (i) the gravitational capture of gas from the surrounding ISM (as a reservoir), or from $\sim{\rm pc}$ to sub-pc scales, as studied in \citetalias{ShiKremerGrudic_2023MNRAS.518.3606S} and resolved in our simulations here; (ii) the mass transfer from the BH accretion disk (as a secondary reservoir) to the event horizon, or from sub-pc to $\sim {\rm km}$ (or $r_{\rm Sch}$) scales, which we \textit{do not} resolve in our simulations. Along with the accretion, gravitational energy is transformed into other forms of energy, like radiation, winds/jets, and cosmic rays. 

At the second phase (smaller scales), not all of the mass/energy flows reach the BH: (i) some fraction of the accretion flow will be powered by kinetic/radiative pressure and form jets/winds; (ii) some fraction of the energy ``leaks'' from the disk and is deposited into the GMC, in the form of radiation and CRs. As a result, intuitively BH feedback mechanisms typically serve as negative factors that suppress the BH accretion in both two ways: (i) reducing the final mass transfer rate onto the BH due to energy and mass loss; (ii) preventing additional gas in GMC from accreting.

Moreover, there is thus a huge mismatch among the scales and physics involved, and the second process is beyond the resolution limit ($\sim 0.1\,{\rm pc}$) of our MHD simulation. So we treat this with a sub-grid model: the BH sink particle accretes gas from the parent simulation on resolved scales, stores the mass inside a gas reservoir (or disk), then transfers the mass to the BH at a rate $\dot{M}_{\rm dep}$, which is also used to calculate mass and energy flow rates for different feedback mechanisms. 

Details of the sub-grid model are expanded below, following the schematic ``order'' of mass flow from the disk to the BH shown in Fig.~\ref{fig:flowchart}. However, we note that the ``order'' is only for defining quantities like the mass loading factor and energy efficiency, not in the chronological sense. \textcolor{black}{Moreover, since the feedback energy efficiency (defined as the ratio between the outflow energy and $\dot{M}_{\rm BH} c^2$) is typically very small (at most $0.1$) in the simulation, we treat $\dot{M}_{\rm BH,0}$ and $\dot{M}_{\rm BH}$ interchangeably in the simulation and this article, without impacting our results at least in the order-of-magnitude sense. For example, the radiation power is defined as $L_{\rm bol} \equiv \epsilon_{\rm r} M_{\rm BH, 0} c^2$ in Fig.~\ref{fig:flowchart}, the relation $L_{\rm bol} \doteq \epsilon_{\rm r} M_{\rm BH} c^2$ still holds as a very good approximation.  }

\subsection{Sub-grid models for BH accretion and feedback}

\subsubsection{Disk mass depletion}

As shown in the flowchart Fig. \ref{fig:flowchart}, gas is captured by sink particles, via resolved gravitational accretion. It is then added to a ``reservoir'' of mass $M_{\rm d}$, which we will call the ``disk'', surrounding the BH of mass $M_{\rm BH}$.
We require a sub-grid model for the accretion rate $\dot M_{\rm d}$ from the disk to the BH, since this determines the feedback properties, so we define $\dot M_{\rm d} = M_{\rm d}/t_{\rm dep}$ in terms fo the depletion time $t_{\rm dep}$.

The classical way to describe the BH accretion disk is the thin disk model \citep[]{ShakuraSunyaev_1973A&A....24..337S}, where the accretion rate follows
\begin{align}
	\dot{M} & \sim  2\pi \alpha \Sigma_{\rm d} c_{\rm s}^2 /\Omega_{\rm K}.\label{equ:alpha-prescription}
\end{align}
Here $\Sigma_{\rm d}$, $c_{\rm s}$, and $\Omega_{\rm K}$ are respectively the surface density, sound speed, and Keplerian frequency at a certain radius of the disk; $\alpha$ is a dimensionless constant that characterizes the effective viscosity. 
Motivated heuristically by Eq.~\eqref{equ:alpha-prescription}, we set
\begin{align}
	t_{\rm dep} = t_{\rm dyn,sink}/(2\pi \alpha).
\end{align}
Note that we redefined $\alpha$ by absorbing the factor of $(\Omega_{\rm K} r_{\rm sink}/c_{\rm s})^2$ and other constants. Here $t_{\rm dyn,sink} \equiv[r_{\rm sink}^3/G(M_{\rm BH}+M_{\rm d})]^{1/2}$ is the dynamical time scale of at the sink radius, which is the fastest possible mass-depletion time. By varying $\alpha$ we may bracket different conditions including slow and fast mass depletion rates.

We caution that for realistic BH accretion, the mass inflow rate can be dependent on the radius since there is mass outflow in the form of winds \citep{BlandfordBegelman_1999MNRAS.303L...1B,HuInayoshiHaiman_2022ApJ...934..132H}. Although in Eq.~\eqref{equ:alpha-prescription} we actually assume a scale-independent mass inflow rate, the effect of BH wind feedback is indeed considered (the next subsection), where we assume only a fraction of the mass can reach the BH while the remaining are deposited back to the GMC in the form of wind and jet.

In the simulation, we also add two constraints to the BH mass transfer. We first set a maximum mass of the accretion disk to represent the effects like fragmentation limiting accretion if mass ``stalls'' in the disk, which is quantified by the upper limit of $M_{\rm d}/M_{\rm BH}$: once $M_{\rm d}$ reaches this limit, no gas cells will be absorbed by the BH in the next time step. Secondly, there is also an upper limit on the Eddington ratio which is defined as $f_{\rm Edd}\equiv \dot{M}_{\rm BH}/\dot{M}_{\rm Edd}$. We vary the upper limit $M_{\rm d}/M_{\rm BH}$. Due to our particular interest in hyper-Eddington accretion which is predicted to be extreme \citep[]{InayoshiVisbalHaiman_2020ARA&A..58...27I}, we fix the upper limit of $f_{\rm Edd}$ to be 1000, but obtained qualitatively similar results so long as this limit is $\gg 1$, as we discuss below.

\begin{figure}
	\centering
	\includegraphics[width=\linewidth]{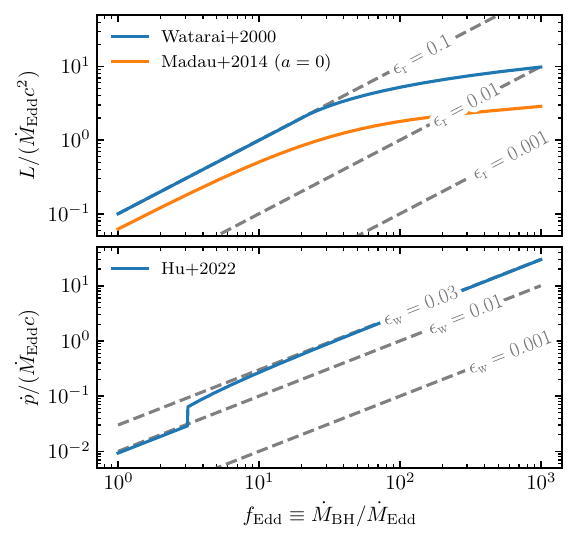}
	\vspace{-20 pt}
	\caption{Sub-grid models for radiative feedback (top panel) and mechanical feedback (bottom two panels). \emph{Top}: energy and momentum ejection rates as a function of accretion rates, including ``log-form'' models inferred from the slim disk (\textit{solid}), and simple ones with constant $\epsilon_{\rm r}\equiv L/(\dot{M}_{\rm BH}c^2)$ (\textit{dashed}). For realistic models the feedback strength  \emph{Bottom}: The momentum ejection rates for mechanical feedback with constant $\epsilon_{\rm w}\equiv \dot E_{\rm kin}/(\dot{M}_{\rm BH}c^2)$ (\textit{dashed}), and the model in \citet{HuInayoshiHaiman_2022ApJ...934..132H} -- \textcolor{black}{though the released kinetic energy is almost constant ($E_{\rm kin }\approx 4\times 10^{-4}\, \dot M_{\rm Edd} c^2 $), the momentum injection rate grows linearly ($\dot p \approx 0.03 f_{\rm Edd} \dot M_{\rm Edd} c$) due to the variable mass loading factor.}}
	\label{fig:feedback-models}
\end{figure}

\begin{table*}
	\centering
	\caption{Free parameters in our BH feedback sub-grid model. Along with the fiducial setup, we also study the impact of different physical processes by varying the corresponding quantities. }
	\begin{tabular}{llll}
		\toprule
		Physical process & Quantity & Fiducial setup & Variations  \\
		\midrule
		\multirow{2}{*}{Disk mass depletion} & $\alpha$ & 0.1 & 0.01, 0.1, 1 \\
		& $\sup (M_{\rm d}/M_{\rm BH})$ & 10 & 1, 10 \\
		\midrule
		\multirow{2}{*}{Radiative feedback} & \textcolor{black}{$\epsilon_{\rm r}\equiv L_{\rm bol}/(\dot{M}_{\rm BH}c^2)$} & Variable: \citet{MadauHaardtDotti_2014ApJ...784L..38M} & 10$^{-9}$ -- 0.1, \citet{WataraiFukueTakeuchi_2000PASJ...52..133W}  \\
		& $Z/Z_\odot$ & 1 & 1,  $10^{-2}$, $10^{-4}$, $10^{-6}$ \\
		
		\midrule
		\multirow{2}{*}{Mechanical feedback} & $\epsilon_{\rm w} \equiv L_{\rm w}/(\dot{M}_{\rm BH,0} c^2)$ & $10^{-6}$ ($10^{-3}$) for $10^{6}\,\rm M_\odot$ ($10^{8}\,\rm M_\odot$) complex & $10^{-8}$ -- $0.1$, \citet{HuInayoshiHaiman_2022ApJ...934..132H} \\
		& $\eta_{\rm w} \equiv \dot{M}_{\rm w}/\dot{M}_{\rm BH,0}$ & 1 & 1, 10, 100 \\
		\midrule
		Cosmic-ray feedback & $\epsilon_{\rm CR}\equiv L_{\rm CR}/\dot{M}_{\rm BH,0}$ & 0 & $10^{-8}$ -- 0.1  \\
		\bottomrule
	\end{tabular}
	\label{tab:parameter_setups}
\end{table*}

\begin{figure*}
	\centering
	\includegraphics[width=.497\linewidth]{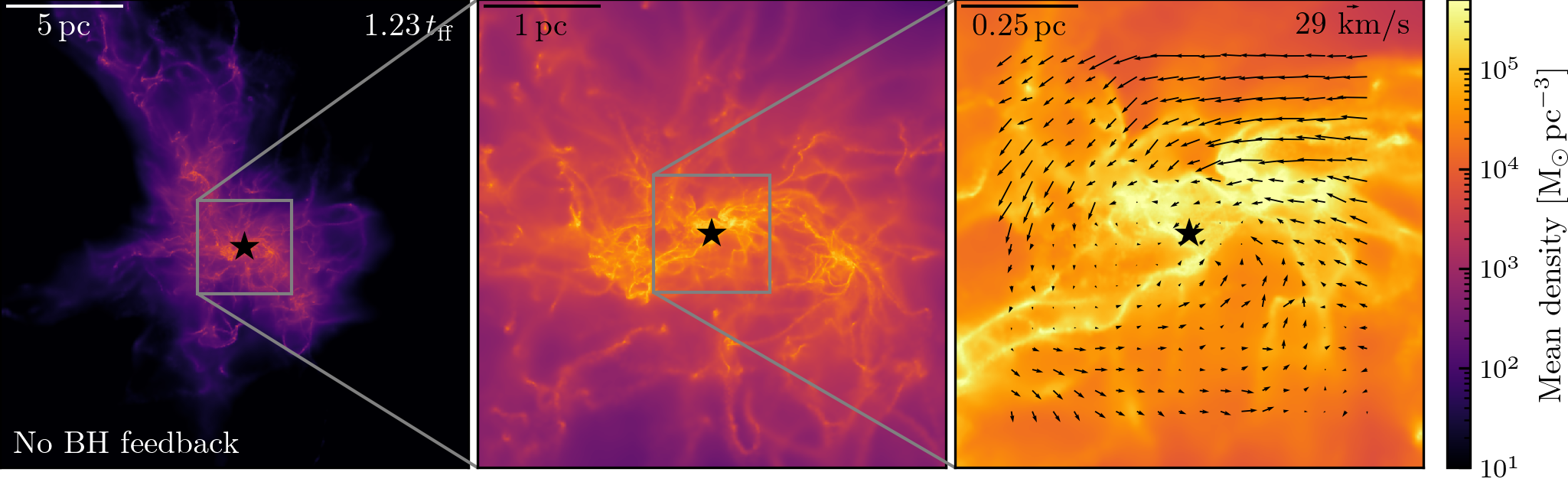}
	\includegraphics[width=.497\linewidth]{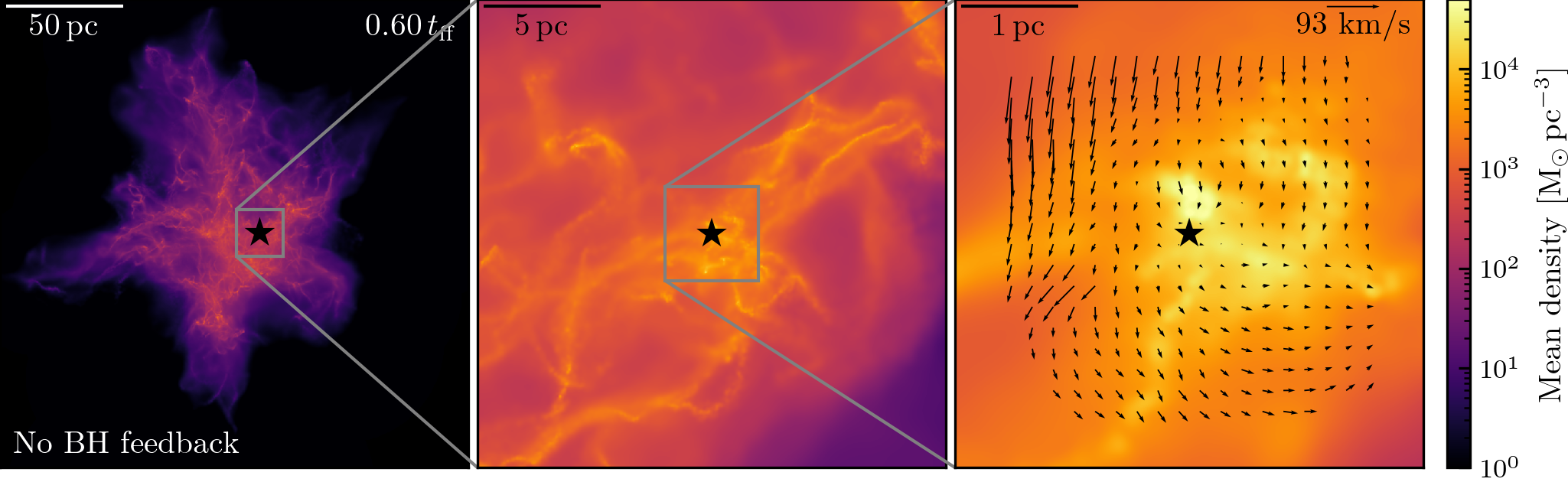}
	\includegraphics[width=.497\linewidth]{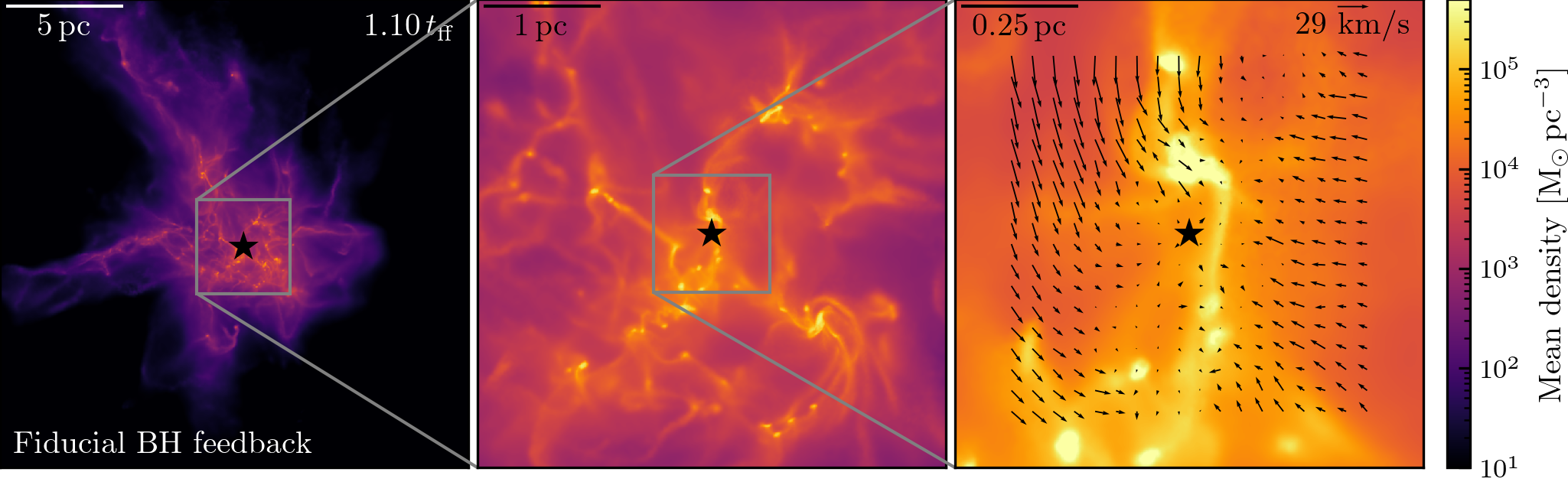}
	\includegraphics[width=.497\linewidth]{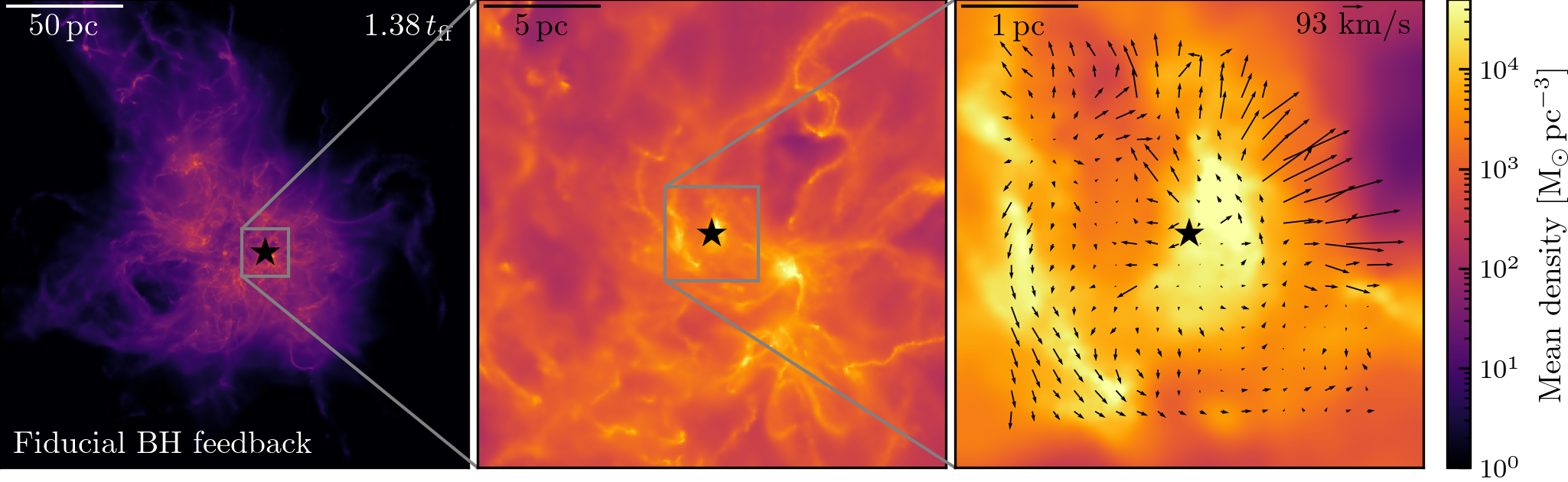}
	\vspace{-5 pt}
	\caption{Visualization of simulations with and without BH feedback physics, each from a snapshot when a BH is undergoing rapid growth. \emph{Left two zoom-in plots}: Gas morphology for the $10^6\,\rm M_\odot$ GMCs without (\textit{top}) or with (\textit{bottom}) BH feedback. From left to right, we zoom in at a BH undergoing hyper-Eddington accretion. As the smallest scale, each panel shows the velocity field near the BH, with the circular velocity ($\sqrt{G M_{\rm cl}/R_{\rm cl}}$) presented as a gauge. 
		\emph{Bottom two zoom-in plots}: the same as the left plots, but for the $10^8\,\rm M_\odot$ GMC.
	}
	\label{fig:visualization}
\end{figure*}

\subsubsection{Mechanical feedback}

In this simulation, we treat the mass and energy flow as shown in Fig.~\ref{fig:flowchart}. For a given $\dot M_{\rm dep}$, we assume some fraction of this will be ejected in the format of winds/outflows/jets, which we combine here as ``mechanical'' outflow and will simply refer to as ``winds'' throughout. We define the jet mass loading factor as $\eta_{\rm w} \equiv \dot{M}_{\rm w}/\dot{M}_{\rm BH,0}$, where $\dot{M}_{\rm w}$ is the jet mass outflow rate, so $\dot{M}_{\rm w} =\dot{M}_{\rm dep}\cdot  \eta_{\rm w}/(1+\eta_{\rm w})$. The kinetic luminosity of the winds can be parameterized with a coefficient $\epsilon_{\rm w}$, such that $L_{\rm w}=\epsilon_{\rm w} \dot{M}_{\rm BH,0}c^2$, or with effective outflow velocity $v_{\rm w}$, then $L_{\rm w}=M_{\rm w} v_{\rm w}^2/2$.

For fiducial simulations, we set $\eta_{\rm w}=1$ and a characteristic $\epsilon_{\rm w}$ such that the mechanical feedback energy is capable of disrupting the whole cloud, i.e., the accumulated jet energy fills the potential well of the GMC: $\int \epsilon_{\rm w} \dot{M}_{\rm BH,0}c^2 \dd t \sim GM_{\rm cl}^2/R_{\rm cl}$. Treating $\epsilon_{\rm w}$ as a constant, we integrate over BH accretion history and find the LHS turns into $\epsilon_{\rm w} \Delta M_{\rm BH,0} c^2$, where $\Delta M_{\rm BH,0}$ is the total mass of gas reaching near the event horizon. We choose a characteristic value that $\Delta M_{\rm BH,c}=10^4\,{\rm M_\odot}$. Then the characteristic $\epsilon_{\rm w}$ is $\epsilon_{\rm w,c} = M_{\rm cl}/\Delta M_{\rm BH,c} (v_{\rm cl}/c)^2$.
Here $v_{\rm cl}^2= GM_{\rm cl}/R_{\rm cl}$ is the characteristic circular velocity of the cloud. Following the energy argument we assumed above, if $\epsilon_{\rm w}\ll \epsilon_{\rm w,c}$ then the mechanical feedback is insignificant; while if $\epsilon_{\rm w} \gg \epsilon_{\rm w,c}$ then the strong mass outflow may disrupt the GMC quickly. It also implies the critical jet velocity at the critical jet luminosity: $v_{\rm w,c}=v_{\rm cl}\sqrt{M_{\rm cl}/\Delta M_{\rm BH,c}}$.

Motivated by simulations of hyper-Eddington accretion \citep{SadowskiLasotaAbramowicz_2016MNRAS.456.3915S}, we assume fiducial $\eta_{\rm w}=1$ but vary $v_{\rm w}$ widely. Moreover, some small-scale simulations fitted sub-grid models which vary $\eta_{\rm w}$ as a function of the accretion rate. In this work, we implemented a version presented in \citet{HuInayoshiHaiman_2022ApJ...934..132H}, \textcolor{black}{which roughly sets $\eta_{\rm w}=f_{\rm Edd} -1$ (where $f_{\rm Edd} \equiv \dot M_{\rm BH}/\dot M_{\rm Edd}$) while keeping a nearly constant kinetic energy output $E_{\rm kin}  \approx 4\times 10^{-4}\,\dot M_{\rm Edd} c^2$ at the super-Eddington regime of $f_{\rm Edd}\gtrsim 3$.}


Numerically, the winds/jets in the simulation are implemented as high-resolution gas cells that are ejected in a bipolar fashion from the BH \citep{TorreyHopkinsFaucher-Giguere_2020MNRAS.497.5292T}. In the rest frame of the BH, these gas cells are ejected along the spin of the BH, which is further determined by the accumulated angular momentum from the gas. A similar implementation was also used and described in \citet{SuHopkinsBryan_2021MNRAS.507..175S,GrudicGuszejnovHopkins_2021MNRAS.506.2199G}.


\subsubsection{Radiative feedback}

Given some remaining accretion rate $\dot{M}_{\rm BH,0}$ (after the loss of winds from $\dot M_{\rm dep}$), we can calculate the bolometric accretion disk luminosity $L_{\rm bol}$ in terms of some radiative efficiency: \textcolor{black}{$L_{\rm bol} = \epsilon_{\rm r}\dot{M}_{\rm Edd}c^2$}. \textcolor{black}{Given $L_{\rm bol}$, we explicitly model the outgoing BH spectrum based on the template in \citet{ShenHopkinsFaucher-Giguere_2020MNRAS.495.3252S} and its effects in our multi-band (X-ray, EUV, FUV, NUV, OIR, far-IR) radiation treatment \citep[as that in][]{HopkinsGrudicWetzel_2020MNRAS.491.3702H}, including its effects on photo-ionization and photo-heating, Compton heating, and radiation pressure on gas and dust, as described in \citet{HopkinsWetzelWheeler_2023MNRAS.519.3154H}}.

Analytical studies and numerical simulations have shown that due to the photon-trapping effect on small scales near BHs, the bolometric luminosity grows slowly with increasing accretion rate (rescaled to $f_{\rm Edd}\equiv \dot{M}_{\rm BH,0}/\dot{M}_{\rm Edd}$), typically in a logarithm form. Due to this reason, super-Eddington accretion is not limited by the radiative energy loss \citep[]{MadauHaardtDotti_2014ApJ...784L..38M}.

In our simulation, we therefore compare various constant-$\epsilon_{\rm r}$ choices with two models for a variable $\epsilon_{\rm r}$ (as shown in Fig.~\ref{fig:feedback-models}). The first one by \cite{WataraiFukueTakeuchi_2000PASJ...52..133W} is analytically derived from the thin disk model, which features a linear form at low accretion rate (with contact $\epsilon_{\rm r}=0.1$) and a logarithm form at high accretion rate. Another form is fitted by \cite{MadauHaardtDotti_2014ApJ...784L..38M} from original simulations of relativistic slim disks in \cite{Sadowski_2009ApJS..183..171S}, which also features dependence on the BH spin parameter $a$. We set $a=0$ for this form in our simulation since \cite{MadauHaardtDotti_2014ApJ...784L..38M} found that the super-Eddington accretion $e$-folding time is nearly independent of $a$. 

\subsubsection{Cosmic rays}

Previous simulations have argued for the importance of cosmic rays (ultra-relativistic particles) accelerated around BHs or jets, e.g., \citet{SijackiPfrommerSpringel_2008MNRAS.387.1403S,GuoMathews_2012ApJ...756..181G,SuHopkinsBryan_2021MNRAS.507..175S}. In addition, CRs also lose energy due to Coulomb interactions, ionization, and catastrophic losses, which also heat the ambient gas \citep[e.g.,][]{GuoOh_2008MNRAS.384..251G}.

In this study, we simulate CRs based on implementations described in \cite{HopkinsSquireButsky_2022MNRAS.509.3779H}, representing them as a ``single-bin'' (relativistic $\sim 1 - 10 \,\rm GeV$ protons) which obey a two-moment streaming+diffusion+acceleration equation, with Alfvenic streaming and a fiducial scattering rate set to $\nu\sim 10^9\,\rm s^{-1}$ calibrated to solar system and Milky Way cosmic ray observations \citep[from Voyager, Fermi, and AMS-02, see][]{HopkinsSquireButsky_2022MNRAS.509.3779H}. All relevant cosmic ray loss/cooling and gas coupling/scattering/ionization/heating details are included as described therein.

In our simulations, CR feedback from BHs is coupled alongside the mechanical feedback (i.e., the newly spawned jet cells). We vary the strength of the BH CR feedback through its energy-loading coefficient $\epsilon_{\rm CR}$, defined by the injection rate $L_{\rm CR}=\epsilon_{\rm CR}\dot{M}_{\rm BH,0} c^2$. Additionally, for each simulation with the BH CR feedback, we set a constant CR energy ejection rate of $10\%$ for supernovae to account for the CR feedback from stars. This particular effect is not considered in other sets of simulations without CRs.

\subsection{Simulation initial conditions}

\begin{figure}
	\centering
	\includegraphics[width=\linewidth]{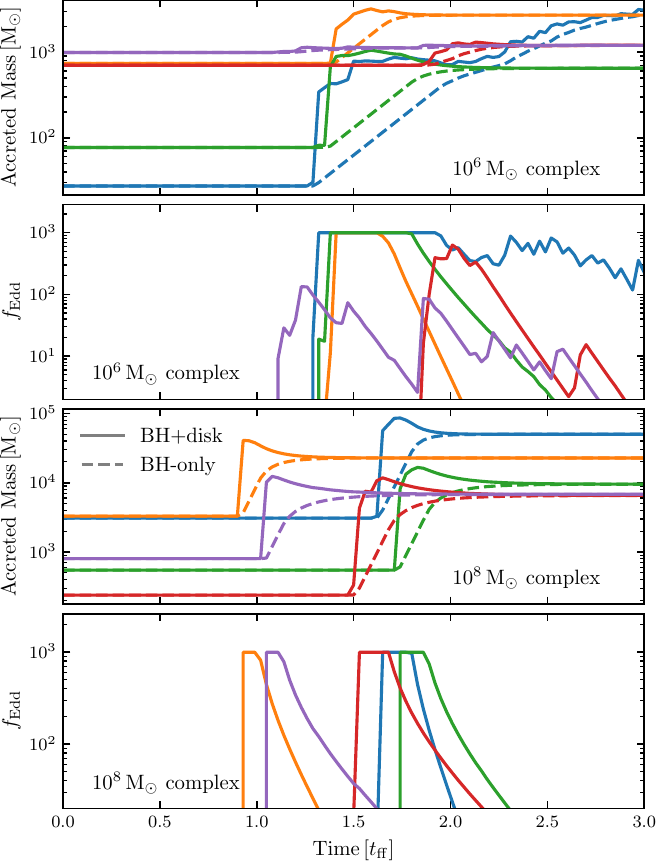}
	\vspace{-15pt}
	\caption{Mass evolution and accretion rate for BH particles in different GMCs. In the first and third panels (from top to bottom), we select 5 BHs with significant mass growth from high-resolution simulations of the two complexes and plot the evolution of the mass bound to the BH (BH+disk, \textit{solid}) and the BH-only mass (\textit{dashed}). In the second and fourth panels, we show the accretion rate for the same BHs, each rescaled to its Eddington accretion rate ($f_{\rm Edd}=\dot M_{\rm BH}/\dot M_{\rm Edd}$).
	}
	\label{fig:mass-evolution}
\end{figure}

We follow the same initial condition (IC) for GMCs as \citetalias{ShiKremerGrudic_2023MNRAS.518.3606S}, but only choose GMCs with the highest initial surface density ($ \Sigma_0 = 13,000\,\rm M_{\odot}/{\rm pc}^2$) with different radii (5 or 50\,pc) and therefore, different masses ($10^6$ or $10^8\,{\rm M_\odot}$). These are chosen as the clouds in \citetalias{ShiKremerGrudic_2023MNRAS.518.3606S} where BHs accrete significantly \textit{without} BH feedback, while other cases would be unlikely to see significant mass growth with the feedback physics. The clouds are initially with the solar metallicity ($Z_\odot$) in the fiducial case. The ICs are listed in Tab.~\ref{tab:gmc_setups}. The time limit for each simulation is $2.5\,{t_{\rm ff}}$, where $t_{\rm ff}=[R_{\rm cl}^3/(8 GM_{\rm cl})]^{1/2}$ is the initial free-fall time of the cloud. For each mass group, there are low (initially with $64^3$ equal-mass gas cells) and high ($128^3$) resolution runs. Force softening is set as described in \citetalias{ShiKremerGrudic_2023MNRAS.518.3606S}.

Each simulation is initialized with a number of BH seeds with random mass, position, and velocity, which are sampled following the same method as \citetalias{ShiKremerGrudic_2023MNRAS.518.3606S}: BHs are randomly distributed following uniform spatial distribution within the volume; the initial velocity magnitude is confined to be below the local circular velocity \citep[for discussions of these choices, see][]{ShiKremerGrudic_2023MNRAS.518.3606S}. In particular, the BH mass follows log-uniform distribution within $10$ ($100$) -- $1000\,{\rm M_\odot}$ ($10^4\,{\rm M_\odot}$) for the $10^6\,{\rm M_\odot}$ ($10^8\,{\rm M_\odot}$) cloud, which covers the range of stellar-mass BHs and IMBHs.

The independent parameters we vary to quantify the impact of BH feedback are listed in Tab.~\ref{tab:parameter_setups}. For each parameter, we set a fiducial value as the ``baseline'' and vary each parameter in turn.

\section{Results}
\label{sec:results}

\subsection{Fiducial results: radiative-inefficient models}

Compared with \citetalias{ShiKremerGrudic_2023MNRAS.518.3606S}, we include a ``log-form'' radiative-inefficient model \citep{MadauHaardtDotti_2014ApJ...784L..38M} to account for the radiative feedback from slim disks and a constant wind energy efficiency. This naturally causes a suppression in the mass accretion rate of seed BHs (as will be described in \S\ref{sec:results:parameters}). Here we focus on the two simulations of $10^6\,\rm M_\odot$ and $10^8\,\rm M_\odot$ cloud complexes with ``high resolution'' (initially with $128^3$ gas cells, see Tab.~\ref{tab:gmc_setups}).

As a first impression, Fig.~\ref{fig:visualization} compares the gas morphology of simulations with or without BH feedback at different scales. From left to right, we zoom in on a BH experiencing significant mass accretion we select. The particular BH which accreted the most mass throughout the simulation time range, and the snapshot shown is also chosen when that BH is in its fastest-accreting phase. Each panel shows the line-of-sight averaged density, which is defined as $\int \rho(s) \dd s/\int \dd s$. At the smallest scales (panels in the right column), we evaluate $\mathbf{v}_{\rm gas} - \mathbf{v}_{\rm BH}$ and show the velocity field within a thin layer of 1/8 of the field of view (centered at the BH) on top of the density field. 

As was also suggested in \citetalias{ShiKremerGrudic_2023MNRAS.518.3606S}, the fast-growing BH seed is often near a dense gas clump ($\rho \gtrsim 10^5 \,\rm M_\odot/pc^3$, typically more than 100 times denser than the mean density). With a small relative velocity $|\mathbf{v}_{\rm gas}-\mathbf{v}_{\rm BH}| < |\mathbf{v}_{\rm BH}|$ which ensures efficient gravitational capture of gas. Despite feedback from BHs, these features are also observed in corresponding simulations of both $10^6\,{\rm M_\odot}$ and $10^{8}\,{\rm M_\odot}$ complexes, which also fit the expectations of the Bondi-Hoyle accretion.

In the absence of BH feedback, the gas in the simulations is relatively dense. Additionally, near the accreting BH without feedback (see the first and third row of Fig.~\ref{fig:visualization}), there are features like disks and spiral arms at $\sim {\rm pc}$ scales at late times, which means that there is coherent gas inflow due to the potential well of the GMC. At this time, there is also star formation in the $10^6\,\rm M_\odot$ GMC near the BH, creating a cavity in the gas distribution (see the first row of Fig.~\ref{fig:visualization}), though not observed in the $10^8\,\rm M_\odot$ GMC through the simulation time. {In contrast, when BH feedback is turned on, additional bubbles, outflows, sheets, and clumps appear and the inflow is relatively incoherent.}

Fig.~\ref{fig:mass-evolution} shows the BH mass growth history in both fiducial simulations, by selecting 5 seed BHs with the highest final-to-initial mass ratios. For each BH, both the BH-only mass and the sink-particle (BH+disk) mass are shown. As in \citetalias{ShiKremerGrudic_2023MNRAS.518.3606S}, single BHs grow rapidly when they encounter dense clumps, while the accretion rate is low most of the time. 

We see little difference in the BH-only mass and the sink particle mass at the early stage of the evolution when the BH accretion rate is low. If the BH feedback is significantly strong, the overall mass accretion is suppressed, and the BH accretion is determined by the material that is already bound to its gravity. Due to the {timescale for} mass transfer from the disk to the BH, there is a ``phase lag'' between $ M_{\rm BH}$ and $M_{\rm sink}$, but the effect is not significant in our fiducial model.

\begin{figure}
	\centering
	\includegraphics[width=\linewidth]{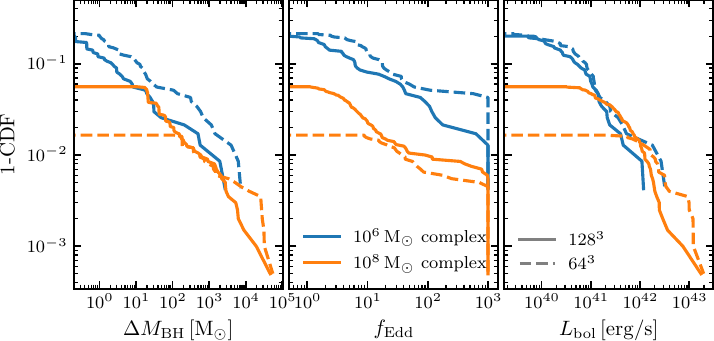}
	\vspace{-15pt}
	\caption{ Cumulative distribution function (CDF) of the accreted mass, Eddington ratio, and bolometric luminosity for  BH populations in simulations with the fiducial BH feedback setups. In the right two panels, we show the distribution of the ``peak'' value throughout the evolution. Both simulations of high (\textit{solid}) and low (\textit{dashed}) resolutions are displayed and show qualitatively similar trends. The artificial flat ``plateau'' at the left of some lines reflects the mass resolution limit of that simulation.
	}
	\label{fig:fiducial-cdf}
\end{figure}

For single BHs, the BH mass accretion rates $\dot{M}_{\rm BH}$ (the actual accretion rate arriving at the BH) are also shown in Fig.~\ref{fig:mass-evolution}. We scale the rate in the unit of the Eddington rate for each BH. Here we show the same 5 BHs selected in Fig.~\ref{fig:mass-evolution}. These samples typically reach hyper-Eddington accretion abruptly at some time in their evolution and maintain the status for a short period of time ($\sim 0.1\,t_{\rm ff}$), during which time $\dot M_{\rm BH}$ is capped by the upper limit set in the code. Once the fast accretion phase is terminated, the mass transfer from the disk to the BH declines with the characteristic scale of $t_{\rm dep}$ as defined in \S~\ref{sec:method}. 

Thus, from Fig.~\ref{fig:mass-evolution}, there is evidence that hyper-Eddington accretion is achievable for BHs in the dense complexes we simulated, even with some BH feedback model. However, the BH feedback can have a negative impact on accretion, which we will quantify in \S~\ref{sec:results:parameters}.

\begin{figure}
	\centering
	\includegraphics[width=\linewidth]{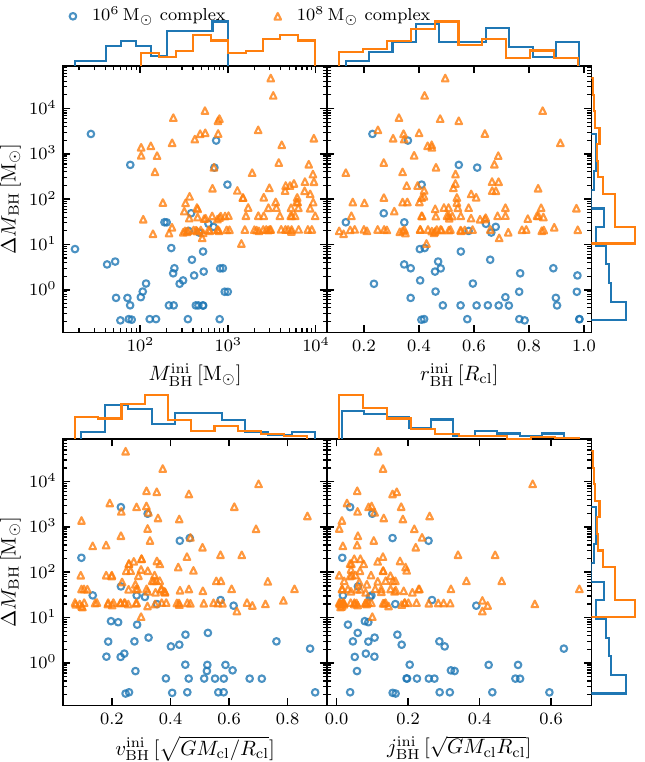}
	\vspace{-15pt}
	\caption{Dependence of BH accretion on initial conditions of BHs. From our high-resolution fiducial simulations, we mark BHs with resolved accretion in a phase space of the accreted mass ($\Delta M_{\rm BH}$) versus The initial mass ($M_{\rm BH}^{\rm ini}$), distance to the cloud's center ($r_{\rm BH}^{\rm ini}$), velocity ($v_{\rm BH}^{\rm ini}$), and specific angular momentum ($j_{\rm BH}^{\rm ini}$). We find only shallow dependence on these initial conditions. We may also clearly see the mass resolution limit from the lower ``cutoff'' in $\Delta M_{\rm BH}$ for the $10^8\,\rm M_\odot$ complex.
	}
	\label{fig:initial-final-mass}
\end{figure}

Fig.~\ref{fig:fiducial-cdf} shows the cumulative distribution function (CDF) of $\Delta M_{\rm BH}$ (relative mass growth at the end of the simulation of each BH) measured from simulations. As expected in \citetalias{ShiKremerGrudic_2023MNRAS.518.3606S}, only a modest fraction ($\lesssim 10\%$) of BH seeds have appreciable mass accretion as the resolution allows. Comparing the two GMCs, we find that BHs in the $10^8\,\rm M_\odot$ GMC have much more significant mass growth, which is again expected from \citetalias{ShiKremerGrudic_2023MNRAS.518.3606S}. Moreover, we compare the low- and high-resolution simulations with the same BH feedback recipes in the same figure and found a qualitatively similar distribution.

Throughout the simulation, we find these BHs with resolved accretion activity typically reach super-Eddington accretion at some stage of the evolution, and a small fraction ($\sim 1\%$) of the total BHs can even reach the preset cap ($f_{\rm Edd}\ge 1000$). We also check the bolometric luminosity, defined as $L_{\rm bol} = \epsilon_{\rm r}(f_{\rm Edd}) \dot{M}_{\rm BH} c^2$. For the $10^6\,\rm M_\odot$ complex, BHs can emit radiation at $\sim 10^{42}\,{\rm erg/s}$ at some stage of their evolution, while the value can be $\sim 10^{42}$ -- $10^{43}\,{\rm erg/s}$ for BHs in the $10^8\,\rm M_\odot$ complex.

\textcolor{black}{For BHs in these two complexes, the CDFs of $\Delta M_{\rm BH}$ and $L_{\rm bol}$ are very similar despite different mass resolutions and number of sampling points (i.e., the number of seed BHs in each simulation). We find that the distribution function of $\Delta M_{\rm BH}$ can be well-fitted with a log-normal distribution: $f(\Delta M_{\rm BH}) \propto \exp[-\log(\Delta M_{\rm BH}/\mu)^2/(2 s^2)]$. For both fiducial runs of the $10^6\,\rm M_\odot$ and $10^8\,\rm M_\odot$ complexes, the best-fit parameters are approximately $\mu = 10^{-2.5}\,\rm M_\odot$, $s=2.25$. Explicitly, the CDF of $\Delta M_{\rm BH}\,[\rm M_\odot]$ is
\begin{align}
    P(\Delta M_{\rm BH} > x) = \frac{1}{2}\left[1-{\rm erf}\left(\frac{\log x +2.5}{2.25\sqrt{2}}\right)\right].
\end{align}
}

In \citetalias{ShiKremerGrudic_2023MNRAS.518.3606S}, we demonstrated that there is weak dependence of the results on the initial properties of BH seeds, like their initial mass, position, and velocity magnitude, as long as they are gravitationally bound to the cloud. Similarly, we explore such dependence for our new simulations by checking $\Delta M_{\rm BH}$ for each BH at the end of the simulation. In each panel of Fig.~\ref{fig:initial-final-mass}, we list BH's initial mass, radii from the GMC's center, velocity magnitude, and the specific angular momentum. Again, for both the $10^{6}\,{\rm M_\odot}$ and $10^8\,{\rm M_\odot}$ cloud, there is no (or little) correlation with $\Delta M_{\rm sink}$. The result means that BH accretion in a turbulent environment is stochastic and much information from the initial condition is smeared in the process. This is why we simulated $N_{\rm BH}\gg 1$ seeds in each cloud -- there is only a $\sim 1 - 10\%$ probability of a seed being ``lucky'' and encountering a dense clump for accretion.

From the same figure, BH accretion has no strong dependence on the initial position or velocity. However, from the last panel, there is stronger evidence that BHs with significant mass growth tend to have low initial specific angular momentum ($\lesssim 0.2\,\sqrt{G M_{\rm cl} R_{\rm cl}}$). This is expected from the evolution of the cloud, since the gravitational collapse brings dense gas inward, and only BH seeds with low specific angular momentum can reach the inner part of the complex.

\subsection{Parameter survey: effects of different physics}
\label{sec:results:parameters}

To check the effects of different physics, we perform a parameter survey by varying parameters in our sub-grid model, as listed in Tab.~\ref{tab:parameter_setups}. For considerations in computational costs, we perform these simulations at low resolution (see Tab.~\ref{tab:gmc_setups}). 

\begin{figure}
	\centering
	\includegraphics[width=\linewidth]{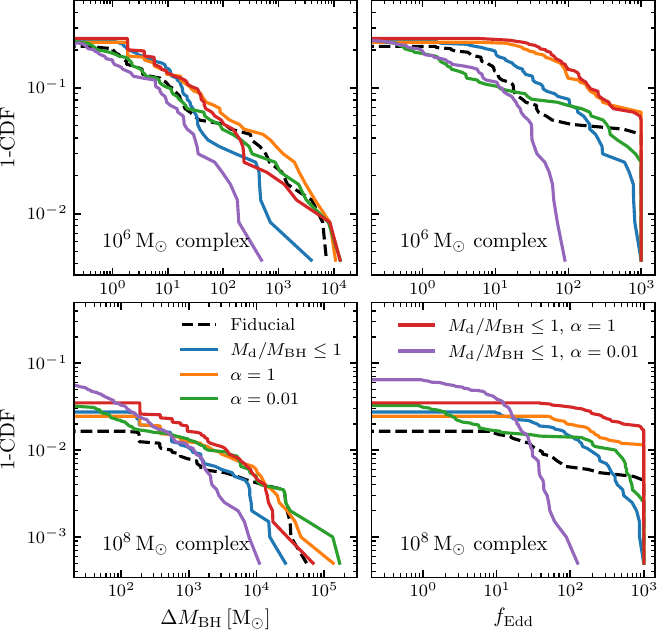}
	\vspace{-15pt}
	\caption{Parameter survey of the accretion model, which utilizes an effective $\alpha$-parameter to quantify the accretion rate. Here we vary $\alpha$ and the upper limit of disk mass. In the fiducial case (\textit{dashed}), $\alpha=0.1$ and $M_{\rm d}/M_{\rm BH} \le 10$.}
	\label{fig:alpha-disk}
\end{figure}

\begin{figure*}
	\centering
	\includegraphics[width=\linewidth]{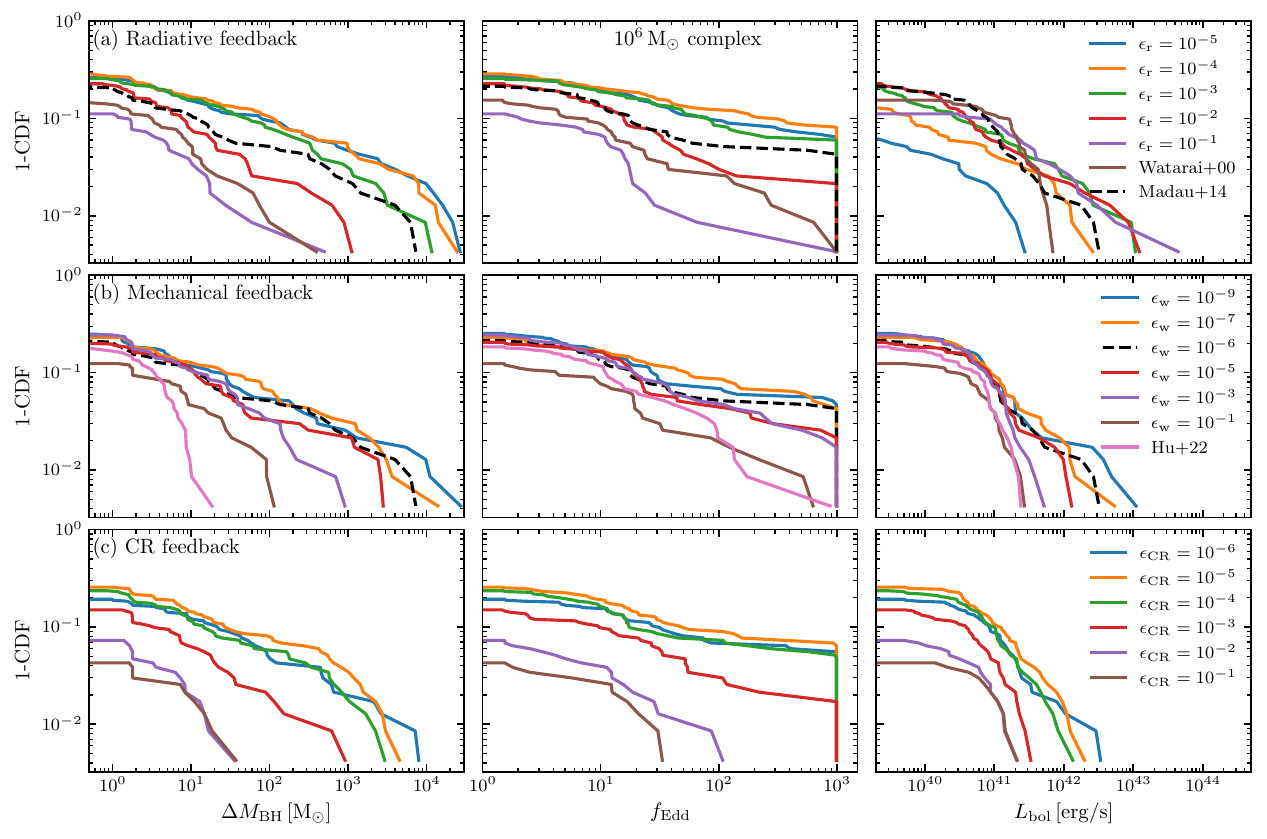}
	\vspace{-20pt}
	\caption{Parameter survey of the energy efficiency in different feedback mechanisms based on experiments with the $10^6\,\rm M_\odot$ complex. From left to right, the three columns show the CDF of accreted mass by BHs, as well as the highest possible $f_{\rm Edd}$ and $L_{\rm bol}$ throughout the evolution. Each row represents a set of simulations with varying models of a specific feedback mechanism (labeled in each left panel), with the fiducial model marked in black dashed lines.
	}
	\label{fig:feedback-cdf-M1e6}
\end{figure*}

\subsubsection{Sub-grid accretion model}

In Fig. \ref{fig:alpha-disk} we show the effect of free parameters in the disk mass depletion: the upper limit of $M_{\rm d}/M_{\rm BH}$, and the effective $\alpha$ parameter for viscosity. In this plot, we show the CDF of the change in BH+disk and the BH-only mass, as well as the CDF of the Eddington ratio for BHs through evolution. Comparing the fiducial case, enforcing a smaller upper limit $M_{\rm d}/M_{\rm BH} \le 1$ leads to accreted mass lower by a factor of $\sim 10$, and fewer BHs can reach hyper-Eddington accretion. This is a natural consequence of the smaller disk mass, which means that less mass can reach the BH+disk system. Because the mass comes in rapidly, it must either be able to accrete extremely rapidly through the disk ($\alpha >1$), or be able to avoid fragmentation and star formation ($M_{\rm d} /M_{\rm BH} >1$) to reach the ``full'' potential growth. But we stress that we still clearly see hyper-Eddington accretion for $M_{\rm d}/M_{\rm BH} \leq 1$.

Then we test higher (1) or lower (0.01) values of the $\alpha$ parameter, in contrast with the fiducial one (0.1). As shown in Eq.~\eqref{equ:alpha-prescription}, smaller $\alpha$ means slower mass transfer from the disk to the BH, this is true in our tests: by fixing $M_{\rm d}/M_{\rm BH}\le 10$, there is a larger deviation between the BH+disk curve and the BH-only curve in the mass CDF once $\alpha$ is smaller, and the Eddington ratio is typically lower for smaller $\alpha$. The trend is also true for tests with fixed $M_{\rm d}/M_{\rm BH}\le 1$.

The parameter $\alpha$ also affects the mass accretion onto the BH+disk system in the simulation. If $\alpha$ is low, the feedback is also weaker due to less mass reaching the BH; if $\alpha$ is high, mass depletion from the disk is efficient and the disk mass can be supplemented quickly in a very gas-rich environment. Both effects are positive factors for accumulating more mass in the BH+disk system. This argument is also reflected in Fig.~\ref{fig:alpha-disk} when comparing tests with different $\alpha$, where $\alpha=1$ and $\alpha=0.01$ tests sometimes have more mass accretion than the fiducial $\alpha=0.1$ case.

Still, we will show below that none of these effects in the sub-grid accretion model is as dominate as the largest variations in the strength of feedback.

\begin{figure*}
	\centering
	\includegraphics[width=\linewidth]{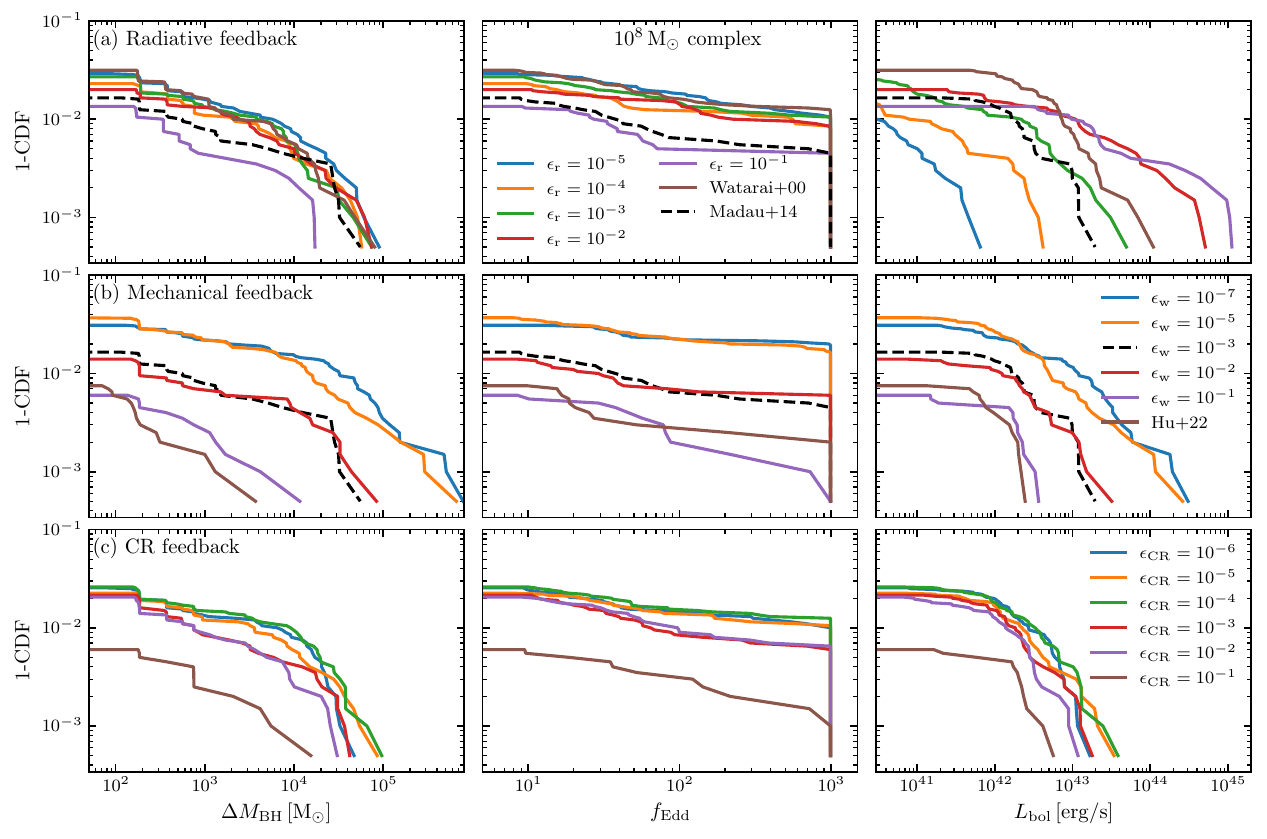}
	\vspace{-20pt}
	\caption{Parameter survey of the energy efficiency in different feedback mechanisms based on experiments with the $10^8\,\rm M_\odot$ cloud complex. Conventions are the same as that in Fig.~\ref{fig:feedback-cdf-M1e6}.
	}
	\label{fig:feedback-cdf-M1e8}
\end{figure*}

\subsubsection{Radiative feedback and metallicity}

We  study the effect of radiative feedback by varying its energy efficiency $\epsilon_{\rm r}$, including two ``log-form'', accretion-rate-dependent models from \cite{WataraiFukueTakeuchi_2000PASJ...52..133W} and \cite{MadauHaardtDotti_2014ApJ...784L..38M}, as well as fixed $\epsilon_{\rm r}$ at $10^{-9}$ -- $0.1$. Similar to the layout of Fig.~\ref{fig:fiducial-cdf}, the top panels of Fig.~\ref{fig:feedback-cdf-M1e6} and Fig.~\ref{fig:feedback-cdf-M1e8} show the CDF of accreted mass, the Eddington ratio, and the bolometric luminosity of the BH population.

Comparing fixed-value energy efficiencies, there is a clear trend that once $\epsilon_{\rm r} \gtrsim 10^{-3}$ ($10^{-1}$) for the $10^6\,\rm M_\odot$ ($10^8\,\rm M_\odot$) GMC, the accretion onto BHs is suppressed. The same behavior appears in the Eddington ratio. Despite relatively strong feedback, there are $1\%$ of BHs reaching the preset cap in the Eddington ratio in many simulations, while the fraction of these BHs drops when the feedback is strong. Finally, the luminosity of BHs typically increases with radiative efficiency as expected.

As also shown in Fig.~\ref{fig:feedback-models}, the radiative efficiency at high accretion rates ($f_{\rm Edd}=1000$) can be as low as $3\times 10^{-2}$ for \cite{MadauHaardtDotti_2014ApJ...784L..38M} and $10^{-2}$ for \cite{WataraiFukueTakeuchi_2000PASJ...52..133W}. Comparing simulations with fixed-value efficiencies, we find that these radiative-inefficient models behave closely like those with fixed $\epsilon_{\rm r}=10^{-3}$ and $\epsilon_{\rm r}=10^{-2}$ respectively. The feedback strength is thus mainly dominated by the hyper-Eddington regime for these models. As the result, the \cite{MadauHaardtDotti_2014ApJ...784L..38M} model is less feedback-dominated.

\begin{figure}
	\centering
	\includegraphics[width=\linewidth]{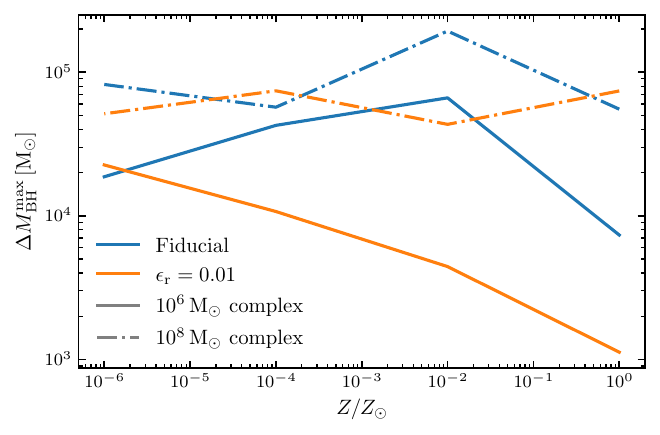}
	\vspace{-20 pt}
	\caption{The impact of the initial metallicity on BH accretion, which is made from a series of simulations varying the initial metallicity ($10^{-6}\,Z_\odot$ -- $Z_{\odot}$) and BH radiative-feedback models (fiducial or $\epsilon_{\rm r}=0.01$).
	}
	\label{fig:metallicity}
\end{figure}

We further study the impact of the GMC's initial metallicity. In the no-BH-feedback limited studied in \citetalias{ShiKremerGrudic_2023MNRAS.518.3606S}, experiments with different initial metallicity found no (or little) correlation between BH accretion and the initial metallicity. However, higher metallicity means tighter coupling between the radiation and wind due to more dust grains in the ISM, which may imply stronger radiative pressure \citep[e.g.,][]{LarsonStarrfield_1971A&A....13..190L}, which is one of the radiative feedback mechanisms (alongside with heating and photoionization) implemented in the simulation. In this study, we vary the initial metallicity for two kinds of radiative models: the fiducial \cite{MadauHaardtDotti_2014ApJ...784L..38M} model and the fixed-value $\epsilon_{\rm r}=0.01$ model (which is in the strong feedback limit). In Fig.~\ref{fig:metallicity}, we plot the maximum accretion in the BH population as a function of the initial metallicity. For most possible combinations of the initial GMC masses and feedback models, there is still no strong dependence on the metallicity. One exception is the $10^{6}\,\rm M_\odot$ GMC with fixed $\epsilon_{\rm r}=0.01$, there is a clear trend that BH accretion drops as the metallicity increases.

This can be explained with the aid of Fig.~\ref{fig:feedback-cdf-M1e6} and Fig.~\ref{fig:feedback-cdf-M1e8}. For the $10^8\,\rm M_\odot$ GMC, both radiative-feedback models above are in the weak-feedback limit even at the solar metallicity (since the CDF of BH mass accretion is close to that with $\epsilon_{\rm r}=10^{-6}$). However, for the $10^6\,\rm M_\odot$ GMC, the $\epsilon_{\rm r}=0.01$ model is beginning to more into a ``strong'' feedback limit as its CDF of mass accretion significantly deviates from the fiducial case. As a result, the GMC metallicity becomes important for that particular case, but in most cases (where we are not very close to this boundary) its effects are weak.

\subsubsection{Mechanical feedback (jets and winds)}
\label{sec:results:mechanical}

Mechanical feedback is also presented in the second rows of Figs.~\ref{fig:feedback-cdf-M1e6} and \ref{fig:feedback-cdf-M1e8}. For models with fixed mass loading factoring $\eta_{\rm w}=1$, we again see the trend that the accretion onto black holes is suppressed once the energy efficiency is sufficiently high, this happens at  $\epsilon_{\rm w}\gtrsim 10^{-5}$ for both complexes. These ``transition points'' happen at lower energy than in the radiative feedback simulations. In terms of $f_{\rm Edd}$, we find the BHs can not reach the maximum value once $\epsilon_{\rm w}\approx 1$. All these simulations assume the radiative feedback efficiency following \cite{MadauHaardtDotti_2014ApJ...784L..38M}, we find the bolometric luminosity decreases when the mechanical feedback is stronger, as expected.

The model from \cite{HuInayoshiHaiman_2022ApJ...934..132H} has variable $\eta_{\rm w}$ and $\epsilon_{\rm w}$. Although its energy efficiency is $10^{-3}$ -- $10^{-2}$ depending on the accretion rate, the accretion rate is highly suppressed: the maximum accretion ($\Delta M_{\rm BH}^{\rm max}$) is even lower than the strong-feedback case with $\epsilon_{\rm w}=0.1$. Despite that the energy efficiency is relatively low, the model has a strong momentum outflow that scales like $\dot{P}\ge 0.01 f_{\rm Edd} \dot{M}_{\rm Edd} c$ (see Fig.~\ref{fig:feedback-models}), resulting very strong BH feedback. Moreover, the fraction of mass flow reaching the BH is low ($1/f_{\rm Edd}$ in the hyper-Eddington phase), making the BH hard to grow. Instead, a much more significant fraction of the material is ejected as mechanical outflow.

\begin{figure}
	\centering
	\includegraphics[width=\linewidth]{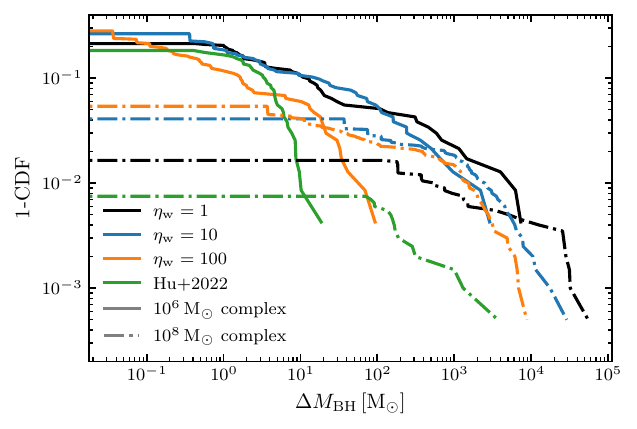}
	\vspace{-20 pt}
	\caption{The effect of jet/wind mass loading factor ($\eta_{\rm w}$) variations, which is made from a series of simulations with constant $\eta_{\rm w}=1, 10, 100$ (but with $\epsilon_{\rm w}$ fixed at the fiducial value), and the variable model in \citet{HuInayoshiHaiman_2022ApJ...934..132H}.
	}
	\label{fig:mass-loading}
\end{figure}

Another important quantity in the mechanical feedback model is the jet loading factor $\eta_{\rm w}$, which determines the fraction of mass flow that goes into the BH ($f_{\rm acc} = 1/(1+\eta_{\rm w})$). We present the result of this experiment in Fig.~\ref{fig:mass-loading}, where we vary $\eta_{\rm w}$ among 1, 10, and 100, corresponding to $f_{\rm acc}$ equals to 0.5, 0.1, and 0.01, while for each variation we keep $\epsilon_{\rm w}$ fixed at the fiducial value of each cloud complex (i.e., change the wind velocity $v_{\rm w}$). For both cloud complexes, we find that the fraction of BHs with mass accretion (with the mass resolution of gas) is almost unchanged despite different $\eta_{\rm w}$, while the maximum accreted mass ($\Delta M_{\rm BH}^{\rm max}$) differs. As a rough estimate from constant-$\epsilon_{\rm w}$ models in Fig.~\ref{fig:mass-loading}, $\Delta M_{\rm BH}^{\rm max} \propto 1/\sqrt{\eta_{\rm w}} \approx \sqrt{ f_{\rm acc}}$. We find that this is in agreement with an analytic scaling that will be covered in \S \ref{sec:discussions}.

\subsubsection{Cosmic rays}

\begin{figure*}
	\centering
	\includegraphics[width=.497\linewidth]{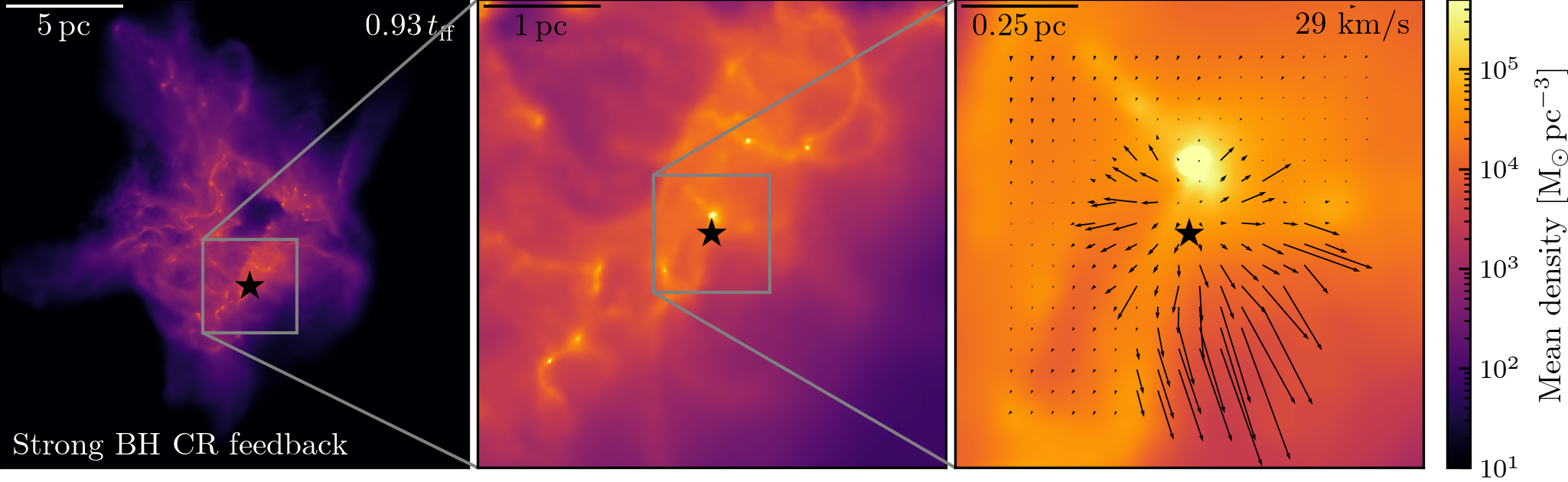}
	\includegraphics[width=.497\linewidth]{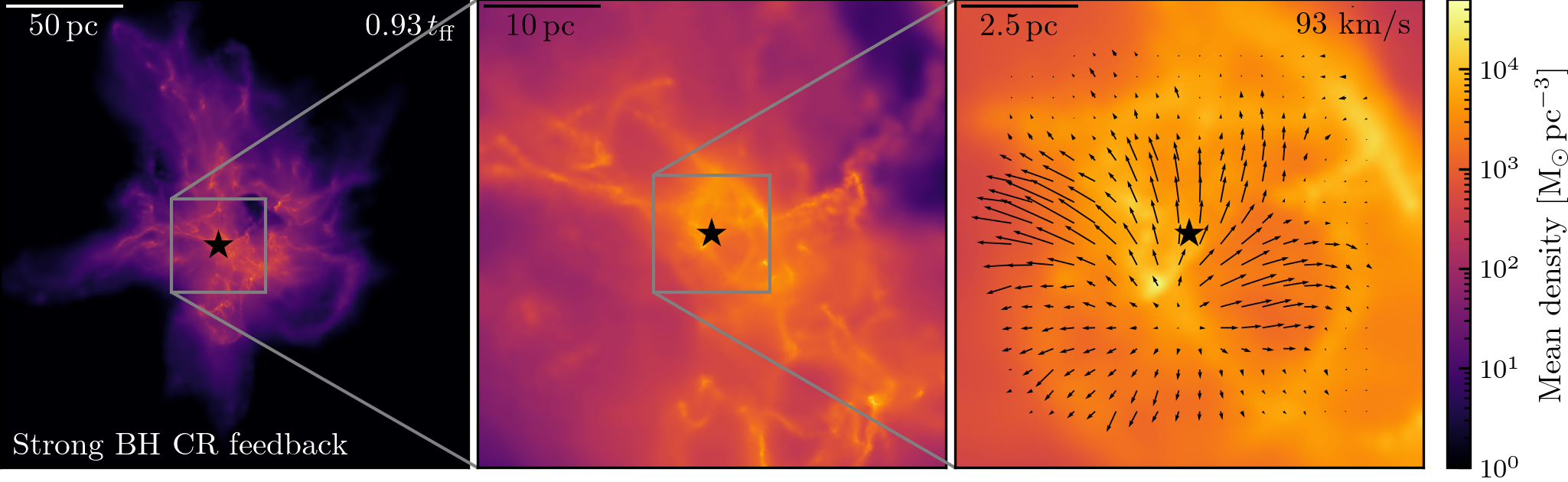}
	\includegraphics[width=.497\linewidth]{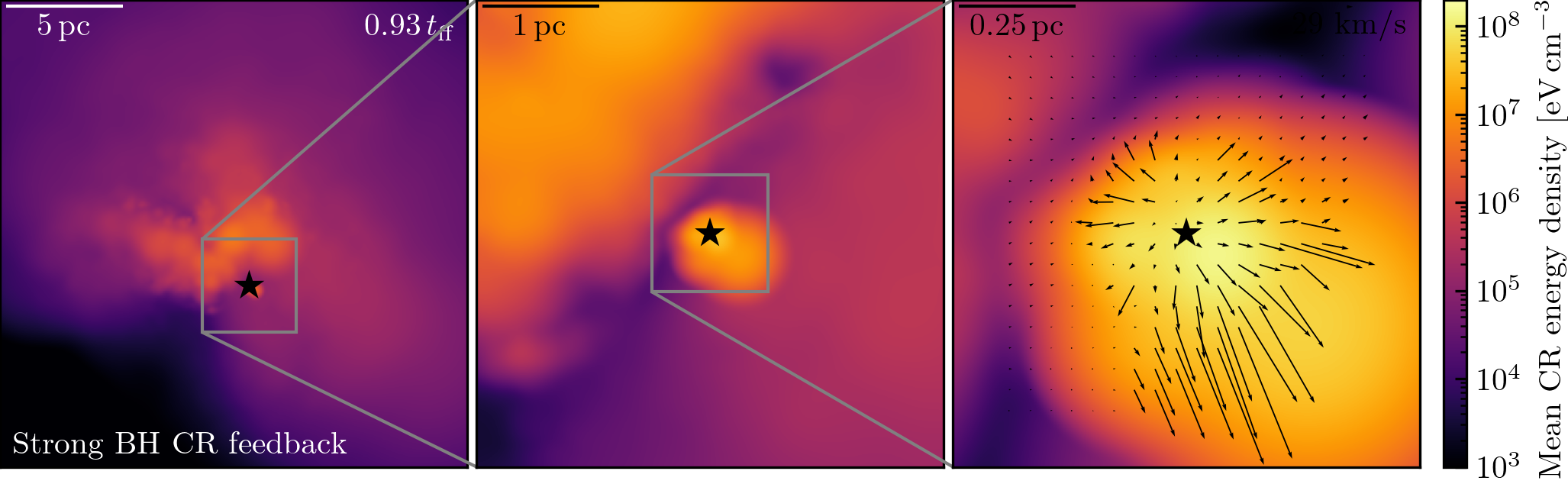}
	\includegraphics[width=.497\linewidth]{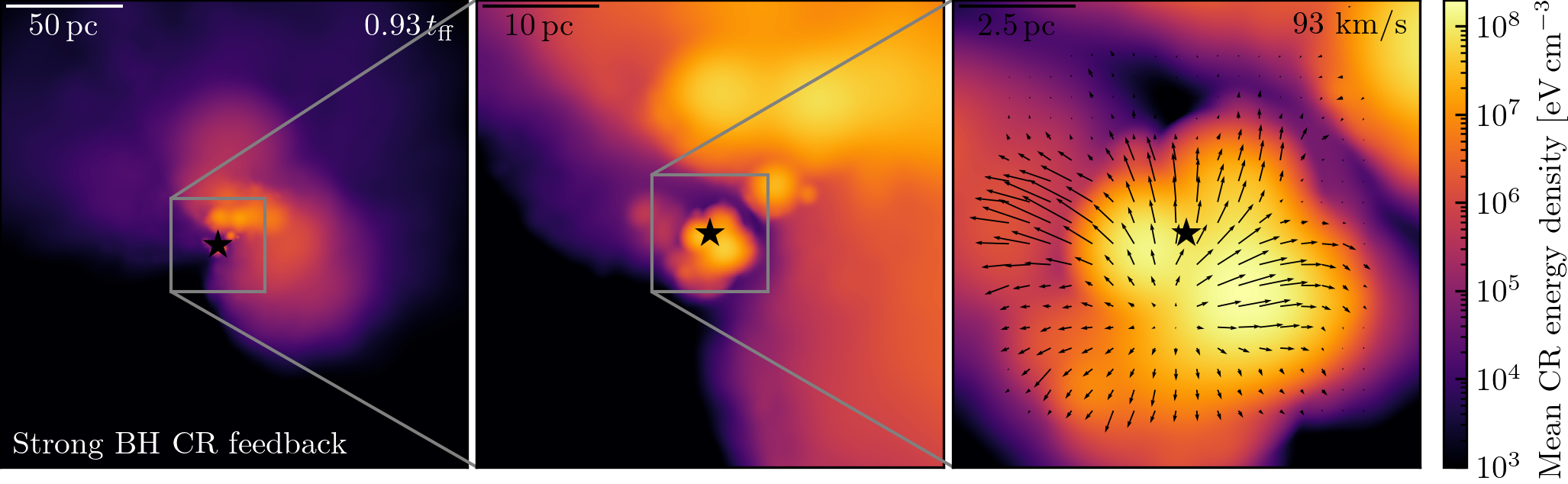}
	\vspace{-5 pt}
	\caption{Visualization of the GMCs at strong cosmic-ray feedback, each from a snapshot with rapid BH growth. \emph{Left two zoom-in plots}: the line-averaged mass density (\textit{top}) and CR energy density (\textit{bottom}) near a BH undergoing runaway accretion, based on a $10^6\,\rm M_\odot$ GMC. We zoom in near the BH, from the left to the right panel, similar to that in Fig.~\ref{fig:visualization}. The line-averaged density is evaluated within the box of each panel. We see a high-energy CR bubble generated by the BH. \emph{Right two zoom-in plots}: the same as the left plots, but embedded in the $10^8\,\rm M_\odot$ GMC. 
	}
	\label{fig:cosmic-ray}
\end{figure*}

Fig.~\ref{fig:cosmic-ray} shows gas morphology and CR energy density for selected simulations with high CR energy efficiency (a $10^6\,\rm M_\odot$ GMC with $\epsilon_{\rm CR}=10^{-3}$ and a $10^8\,\rm M_\odot$ GMC with $\epsilon_{\rm CR}=10^{-2}$), and zooms in toward a rapidly-accreting BH. As also suggested in Fig.~\ref{fig:visualization}, the BH is located at a dense clump in the GMC. Due to strong CR feedback, there is also strong outflow near the BH. Due to CR feedback, there is a high-energy CR ``bubble'' near the BH, whose energy density is comparable with CRs generated through stellar feedback. However, these bubbles are of the size of $\sim 0.25\,\rm pc$ to $\sim 2.5\,\rm pc$, significantly smaller than the scale of stellar CRs (typically $\sim R_{\rm cl}$). As a result, there is a higher energy (pressure) gradient from the high-energy CR bubbles near the BH, which may then expel gaseous material from the BH or even disrupt the whole GMC.

From Figs.~\ref{fig:feedback-cdf-M1e6} and \ref{fig:feedback-cdf-M1e8} we find that BH CR feedback can be important for BH accretion in these dense GMCs we simulated if the energy efficiency is high. For example, BH accretion is significantly suppressed when $\epsilon_{\rm CR} \gtrsim 10^{-3}$ for the $10^6\,\rm M_\odot$ GMC, or $\epsilon_{\rm CR}\gtrsim 10^{-1}$ for the $10^8\,{\rm M_\odot}$ GMC. We also see a similar trend in $f_{\rm Edd}$ and $L_{\rm bol}$. Despite that CR feedback from SNe is also included in this set of simulations, we find no clear difference when comparing the $\epsilon_{\rm CR}=10^{-6}$ run and the run with fiducial setups (without SNe CR feedback), which means this particular stellar feedback mechanism does not bring new effects to BH accretion/feedback.


\section{Discussion}
\label{sec:discussions}

\subsection{BH Feedback and BH accretion}

For all the feedback mechanisms simulated, we see (as expected) that sufficiently strong BH feedback suppresses BH accretion. To further quantify the effect of feedback, we show the relationship between the feedback strength (indicated by energy efficiencies $\epsilon_{\rm r}$, $\epsilon_{\rm w}$, and $\epsilon_{\rm CR}$) and the maximum of the accreted mass of the BH population in Fig. \ref{fig:feedback-scaling}. Besides the fixed-value models, we also show some specific sub-grid models for reference, in which energy efficiency is estimated at the hyper-Eddington regime (at $f_{\rm Edd}=1000$). 

For all mechanisms, there is a plateau at the weak-feedback (low energy-efficiency) end, which simply represents the regime where that particular feedback mechanism does \textit{not} regulate BH growth significantly. However, when feedback is stronger, there is a drop in $\Delta M_{\rm BH}^{\rm max}$. We find that this feedback-regulated limit can be approximated with simple analytic arguments at the order-of-magnitude level.

\subsubsection{The feedback-regulated limit: scalings}

We consider a simple self-regulated model akin to e.g., \citet{SilkRees_1998A&A...331L...1S} or \citet{MurrayQuataertThompson_2005ApJ...618..569M,OstrikerChoiCiotti_2010ApJ...722..642O}. First, consider just radiative and mechanical feedback, since the fiducial model neglects CRs. If accretion is momentum-limited, then BH growth will cease when the ``feedback force'' $F_{\rm feedback} = \dot p_{\rm feedback}$ (momentum injection rate due to feedback) equals the gravitational force on gas in the parent complex, $F_{\rm feedback} \sim F_{\rm grav} \sim G M_{\rm cl}^2/R_{\rm cl}^2$. 

For radiative feedback, the radiation pressure is dominated by single-scattering (with at most $\mathcal{O}(1)$ correction from IR multiple scattering) even in our dense clouds, so $\dot p_{\rm rad} \sim L /c \sim \epsilon_{\rm r} \dot M_{\rm BH} c$. Similarly, for mechanical feedback, $\dot p_{\rm mech} \sim \dot M_{w} v_{\rm w} \sim \sqrt{2\eta_{\rm w} \epsilon_{\rm w}} \dot M_{\rm BH} c$.

For CRs, the momentum injection is somewhat less straightforward because it arises via partial confinement/scattering of CRs generating a CR pressure gradient which accelerates the gas. From the simple analytic spherically-symmetric CR-pressure-driven wind solutions in \citet{JiChanHummels_2020MNRAS.496.4221J,HopkinsSquireButsky_2022MNRAS.509.3779H}, we can approximate this around a dominant point source as $\dot{p}_{\rm CR} \sim \int d^{3}{\bf x}\,\hat{\bf r} \cdot \nabla P_{\rm CR} \sim \ell^{2}\,P_{\rm CR} \sim \dot{E}_{\rm cr}/\bar{v}_{\rm stream,eff}$, where $\dot{E}_{\rm CR} \sim \epsilon_{\rm CR}\,\dot{M}_{\rm BH}\,c^{2}$ and $\bar{v}_{\rm stream,eff}$ is some effective bulk CR streaming speed\footnote{This is just related to the usual spherical wind solution, $u = \dot{E}/4\pi\,v_{\rm eff}\,r^{2}$.} ($\sim \kappa/3\,\ell$ in terms of some characteristic CR gradient scale-length $\ell$, in the diffusion-dominated regime). So $\dot{p}_{\rm CR} \sim (\epsilon_{\rm CR}\,\alpha\,c/\bar{v}_{\rm stream,eff})\,\dot{M}_{\rm BH}\,c$ where $\alpha$ collects the order-unity uncertainties in the scalings above. Inserting our assumed $\kappa$ and assuming $\ell \sim R_{\rm cl}$, or checking the CR properties in the simulations directly, we see the CRs are not strongly confined, so the prefactor $\alpha\,c/\bar{v}_{\rm stream,eff} \sim 10$, roughly. 

So taking $\dot{p}_{\rm feedback} = \dot{p}_{\rm rad} + \dot{p}_{\rm mech} + \dot{p}_{\rm CR}$, we find $\dot p_{\rm feedback} \sim (\epsilon_{\rm r} + \sqrt{2\,\eta_{\rm w}\,\epsilon_{\rm w}} + 10\,\epsilon_{\rm CR}) \dot M_{\rm BH} c$. Equating this feedback force to the self-gravity of the cloud complex will thus determine the maximum $\dot M_{\rm BH}$ possible. From Fig.~\ref{fig:mass-evolution}, we see the BH growth is dominated by phases with a high accretion rate, which typically reaches $f_{\rm Edd} \sim 1000$. The peak accretion rate can then be parameterized as $\dot M_{\rm BH} \sim f_{\rm Edd} M_{\rm BH}/t_{\rm Sal}$. Note that $\Delta M_{\rm BH}^{\rm max} \sim M_{\rm BH}^{\rm max}$ is reached for some BHs with the most significant mass growth, we have
\begin{align}
	& \Delta M_{\rm BH}^{\rm max}  \sim M_{\rm BH}^{\rm max} \sim \frac{G M_{\rm cl}^2}{R_{\rm cl}^2} \frac{t_{\rm Sal}}{f_{\rm Edd} c} \frac{1}{\epsilon_{\rm r} + \sqrt{2\,\eta_{\rm w}\epsilon_{\rm w}}+ 10\, \epsilon_{\rm CR}} \label{equ:fitting-relation}\\
	& \sim 3\times 10^4 \,{\rm M_\odot}\, \left(\frac{M_{\rm cl}}{10^6\,{\rm M_\odot}}\right)^2 \left(\frac{R_{\rm cl}}{5\,{\rm pc}} \right)^{-2} \frac{1}{f_{\rm Edd}(\epsilon_{\rm r} + \sqrt{2\,\eta_{\rm w} \epsilon_{\rm w}}+10\,\epsilon_{\rm CR})}. \nonumber
\end{align}

\begin{figure}
	\centering
	\includegraphics[width=\linewidth]{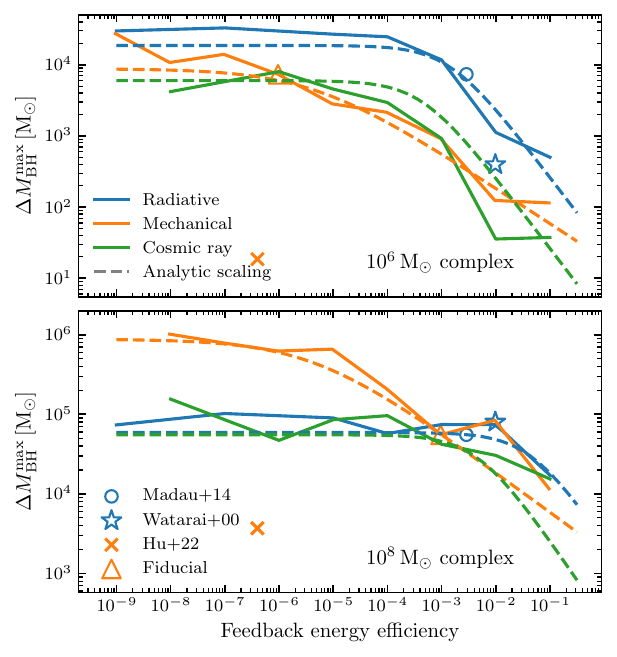}
	\vspace{-20 pt}
	\caption{ Dependence of BH accretion on the energy efficiency of specific feedback mechanisms for the $10^6\,\rm M_\odot$ (\textit{top panel}) and $10^8\,\rm M_\odot$ (\textit{bottom panel}) GMCs. We summarize experiments with fixed-value energy efficiencies (\textit{solid}) and show the scaling relation (\textit{dashed}) based on Eq.~\eqref{equ:fitting-relation}. Several more sophisticated sub-grid models are also displayed for reference, for each of them the energy efficiency is calculated at the hyper-Eddington regime of $f_{\rm Edd}=1000$. 
	}
	\label{fig:feedback-scaling}
\end{figure}

With Eq.~\eqref{equ:fitting-relation}, we plot the scaling relations for each feedback mechanism variation in Fig.~\ref{fig:feedback-scaling}. First, we check models varying radiative and mechanical feedback efficiency, i.e., $\epsilon_{\rm CR}=0$. As described previously, for the fiducial mechanical feedback model, we input $\epsilon_{\rm w}=10^{-6}$ for the $10^{6}\,\rm M_\odot$ complex and $\epsilon_{\rm w}=10^{-3}$ for the $10^{8}\,\rm M_\odot$ one; for the fiducial radiative feedback, we input $\epsilon_{\rm r} = 3\times 10^{-3}$ (c.f. Fig.~\ref{fig:feedback-models}) since $f_{\rm Edd}=1000$ is reached. Varying the feedback efficiency, we find both scaling relations for the radiative and mechanical feedback are in good agreement with the simulations.

Considering CRs, we also find the scaling relation agrees quite well with the simulations, though from our analytic scaling, we see that this is more sensitive to other uncertainties besides $\epsilon_{\rm CR}$, namely CR transport physics. It remains deeply theoretically uncertain how well-coupled GeV CRs are in dense, neutral-phase ISM gas \citep[see e.g.][]{Zweibel_2017PhPl...24e5402Z,KrumholzCrockerXu_2020MNRAS.493.2817K,HopkinsChanJi_2021MNRAS.501.3640H}, and the CR transport model we adopted is purely phenomenological, so the details of this scaling should be taken with more caution. Nonetheless, it provides some order-of-magnitude guidance.

\subsection{BH feedback and star formation}

\begin{figure}
	\centering
	\includegraphics[width=\linewidth]{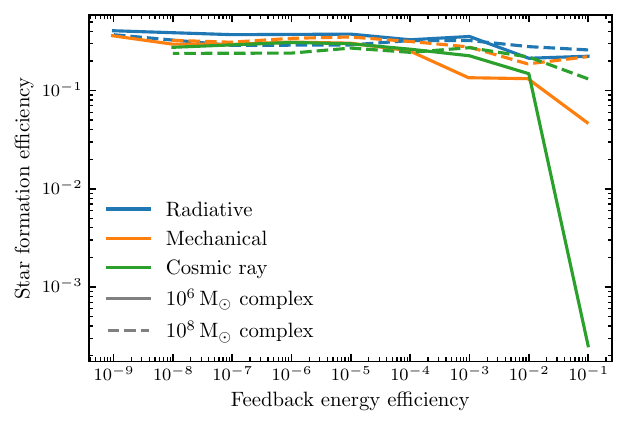}
	\vspace{-20 pt}
	\caption{The star formation efficiency (SFE) in our experiments with various fixed-value BH-feedback energy efficiencies. For the $10^6\,{\rm M_\odot}$ GMC, there is (sometimes extreme) suppression of the SFE at high energy efficiency.
	}
	\label{fig:feedback-sfe}
\end{figure}

Fig.~\ref{fig:feedback-sfe} shows the star formation efficiency (SFE, here defined as the ratio between the stellar mass at the end of the simulation and the initial total gas mass) for different kinds of feedback mechanisms and various energy efficiencies. When the energy efficiency is sufficiently low, we find that in both the $10^6\,\rm M_\odot$ and $10^8\,\rm M_\odot$ GMCs, the SFE is $\sim 0.3$. This is expected since the two GMCs have the same initial mean surface density \citep[][and more citations in \S~\ref{sec:introduction}]{GrudicGuszejnovHopkins_2018MNRAS.481..688G}. Then once the energy efficiency increase beyond $\sim 10^{-3}$, the SFE for the $10^6\,\rm M_\odot$ GMC drops below $0.1$. However, the impact on the $10^8\,\rm M_\odot$ is not significant.

A very special case is the $\epsilon_{\rm CR}=0.1$ run, in which the SFE drops to a very low value ($<10^{-3}$). With further inspection of the run, we find that the cloud was disrupted by strong CR winds at $\sim 0.3\,t_{\rm ff}$, which is much shorter than the characteristic time scale of star formation at the low-BH-feedback limit (from simulations it is $t^{\rm SF}_{1/2}\sim 1.25\,t_{\rm ff}$, defined as the time when the SFE reaches half of the final value). As a result, the SFE almost freezes at the (extremely low ) star-formation level at the time of GMC's disruption.

We may roughly explain this quantitatively. From our simulations, a typical BH growth history is dominated by a few ``bursty'' super-Eddington accretions which happen in a short period of time (typically less than $0.05\,t_{\rm ff}$). At a time scale much longer than the bursty growth, the BH mass is then anticipated to be $M_{\rm BH}\sim M_{\rm BH}^{\rm ini}\exp(\langle f_{\rm Edd} \rangle t/t_{\rm Sal})$. Applying this to the BH with the most significant mass growth, we find the disruption time of the GMC due to strong BH feedback to be 
\begin{align}
	t_{\rm d} \sim \frac{t_{\rm Sal}}{\langle f_{\rm Edd} \rangle} \ln \left(1+ \frac{\Delta M_{\rm BH}^{\rm max}}{M_{\rm BH}^{\rm ini}}\right).
\end{align}
For each simulation, $M_{\rm BH}^{\rm max}$ and $\langle f_{\rm Edd} \rangle$ are presented in the middle columns of Figs.~\ref{fig:fiducial-cdf}, \ref{fig:feedback-cdf-M1e6}, and \ref{fig:feedback-cdf-M1e8}; and $M_{\rm BH}^{\rm max}$ (or $\Delta M_{\rm BH}^{\rm max}$) is also analytically evaluated from Eq.~\eqref{equ:fitting-relation}. In general, we find that $\langle f_{\rm Edd} \rangle$ is approximately 10 -- 100, and there is $M_{\rm BH}^{\rm max} \gg M_{\rm BH}^{\rm ini}$ so $t_{\rm d}$ is generally much larger than the free-fall time of the two GMCs we simulated. However, when there is very strong BH feedback, $\Delta M_{\rm BH}^{\rm max}$ is low. Thus there are chances that $\Delta M_{\rm BH}^{\rm max} \ll M_{\rm BH}^{\rm ini}$ and $t_{\rm d} \sim t_{\rm Sal} \Delta M_{\rm BH}^{\rm max}/(\langle f_{\rm Edd} \rangle M_{\rm BH}^{\rm ini})$ is smaller than (or comparable  with) the free-fall time. For the $\epsilon_{\rm CR}=0.1$ run of the $10^6\,{\rm M_\odot}$ complex, we measure and find $M_{\rm BH}^{\rm ini}=702\,\rm M_\odot$ and $\Delta M_{\rm BH}^{\rm max}=740\,\rm M_\odot$ (thus $\Delta M_{\rm BH}^{\rm max}/M_{\rm BH}^{\rm ini} \approx 0.05$), resulting $t_{\rm d}\sim 2/\langle f_{\rm Edd} \rangle\,{\rm Myr}$. The time scale is comparable to (or substantially smaller than if $\langle f_{\rm Edd} \rangle$ is large) the free-fall time of the GMC ($0.19\,\rm Myr$). 

We also note that the argument is dependent on the initial mass of the BH, thus $t_{\rm d}$ is essentially short only when the ``lucky'' seed BH is initially massive. As a result, the extreme suppression of SFE does not always happen even if the energy efficiency is high, which is true for most experiments in Fig.~\ref{fig:feedback-sfe} with high energy efficiency. 

In summary, we find that BH feedback will not affect the star formation efficiency unless it is in a highly feedback-dominated/regulated regime and relatively lower-mass cloud complex with high mass ``active'' seeds  ($M_{\rm BH} \gtrsim 10^3\,\rm M_\odot$ with $\epsilon \gtrsim 10^{-3}$ for the $10^6\,\rm M_\odot$ GMC). The significance of the impact is largely decided by how soon the winds induced by BH feedback ``interrupt'' star formation (which happens at the free-fall time scale) -- in some particular models (e.g., very strong feedback from a massive BH seed), even if BH accretion is limited, the strong BH feedback is still able to disrupt the whole GMC at the very early stage of the star-formation history and result in very low star formation efficiency.

\subsection{Caveats and outlook}

Because of the huge mismatch in scales of the BH accretion/feedback problem (from kpc/pc scales to the Schwarzschild radius), it is difficult to simulate every aspect in great detail. In this study, we concentrate on larger scales, where the gas inflow towards BHs' gravitational capture radii is resolved but dynamics of small scales (gas cells that are considered ``gravitationally bounded to the BH'' including, of course, the accretion disks themselves) are described by sub-grid models with a few free parameters (as listed in Tab. \ref{tab:parameter_setups}). However, the setup is sufficient to answer the question we were asking: whether there could be enough fuel from large scales (kpc/pc) reaching small scales (sub-pc) to fuel rapid BH growth in the presence of BH feedback. The setup is thus also able to study the impact of different sub-grid parameters on BH accretion, as an attempt to bridge the dynamics of small and large scales.

Still, the sub-grid models and resolution of our simulations do not allow us to connect the large and small scales in great self-consistency. For example, the energy efficiencies of different feedback mechanisms are treated as arbitrary inputs which are not dependent on BH ambient boundary conditions like magnetic field and angular momentum flow. There is thus plenty of space for improvement.

Recent zoom-in simulations of BH accretion or feedback may shed light on this problem \citep[e.g.,][]{TalbotBourneSijacki_2021MNRAS.504.3619T,GuoStoneKim_2023ApJ...946...26G,HopkinsSquireSu_2024OJAp....7E..19H,HopkinsSquireQuataert_2024OJAp....7E..20H}. With the technique of super-Lagrangian refinement, small-scale structures (e.g., BH accretion disks or jets) and dynamics (e.g., the magnetic field in disks, jet launching mechanisms, BH-disk interactions) may also be simulated with higher level of self-consistency.

Finally, as also described in \citetalias{ShiKremerGrudic_2023MNRAS.518.3606S}, next-generation simulations which resolve \textit{individual} stars \citep[]{GrudicGuszejnovHopkins_2021MNRAS.506.2199G,GuszejnovGrudicHopkins_2021MNRAS.502.3646G} may enable other important processes like BH seed formation, stellar merging, and BH accretion/feedback in the same simulation.

\section{Conclusions}
\label{sec:conclusions}
\begin{figure}
	\centering
	\includegraphics[width=\linewidth]{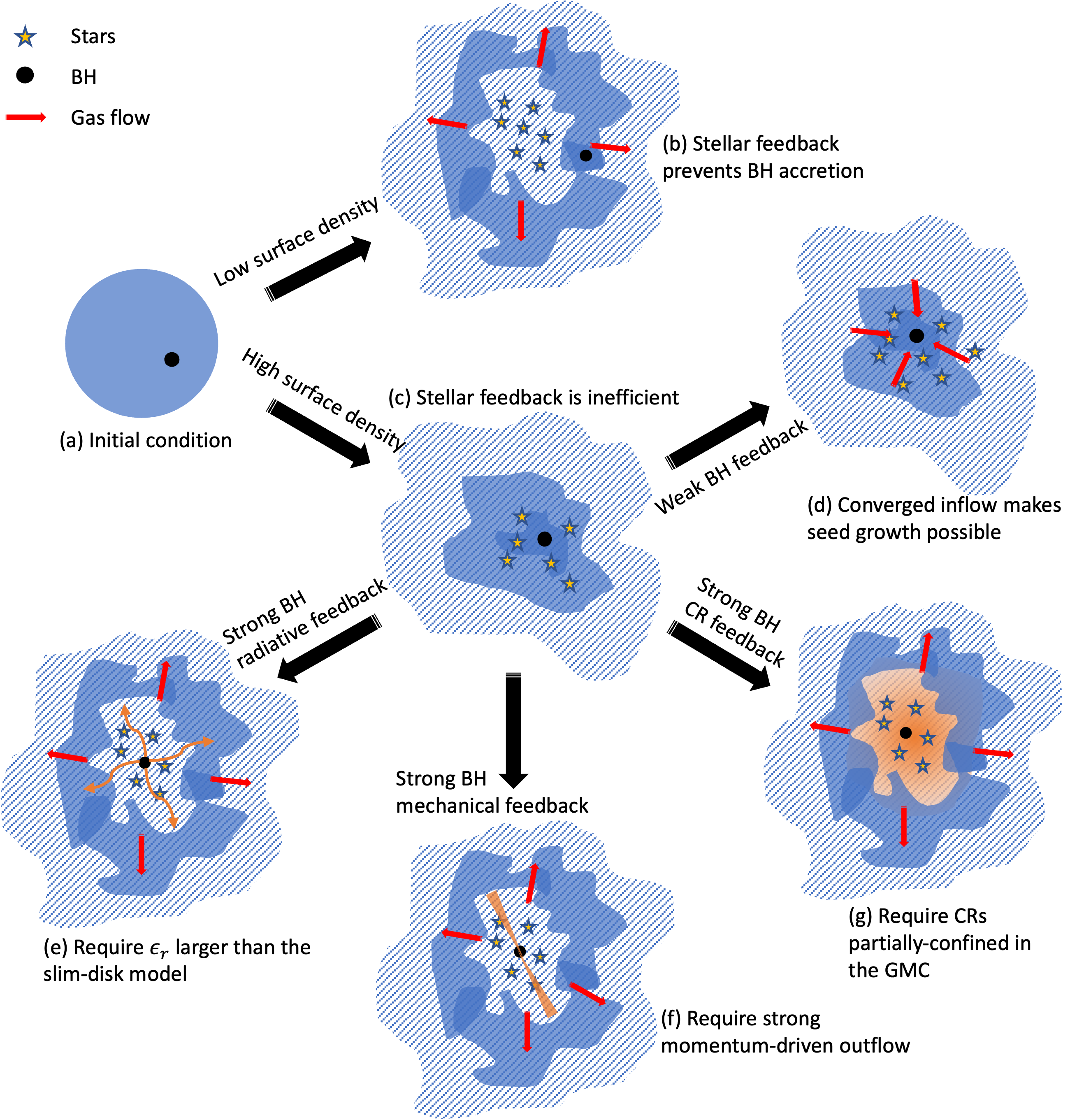}
	\vspace{-10 pt}
	\caption{ Overview of seed BH accretion and feedback in star-forming GMCs. (a). Initial condition, where the seed BH is inside a pre-star-formation, turbulent GMC. (b). If the self-gravity of the GMC (characterized by its mean surface density), star formation and feedback dominate and disrupt the GMC, suppressing BH accretion. (c). If the GMC's self-gravity is strong, gas is kept bound, creating a proper environment for BH accretion and feedback. (d). If the BH feedback is weak, there is steady accretion flow towards the BH if it is at the center of the potential well. (e). (f). (g). Strong feedback in different forms may suppress both BH accretion and star formation, this process is dominated by momentum outflow from BHs and can be approximated with the analytic argument as in Eq.~\eqref{equ:fitting-relation}.
	}
	\label{fig:schemetic}
\end{figure}

This study is a parameter-space survey of seed BH accretion and feedback in star-forming GMCs, with key parameters listed in Tab.~\ref{tab:parameter_setups}. We focus on high surface-density clouds which are ideal environments for runaway BH accretion in the weak BH feedback limit \citepalias{ShiKremerGrudic_2023MNRAS.518.3606S}, and find that BH feedback self-regualtes BH growth. Our major conclusions are listed below and illustrated in Fig.~\ref{fig:schemetic}.

\begin{enumerate}
	\item Even with feedback, the BH accretion scenario is the same as that presented in \citetalias{ShiKremerGrudic_2023MNRAS.518.3606S}. Rapid BH accretion happens when BHs semi-randomly encounter dense gas clumps with low relative velocities. This happens only when the dynamics of both BH and gas are dominated by GMC's self-gravity, i.e., when the cloud has high enough surface density that stellar feedback is inefficient, and the BH has low enough mass that BH feedback is (initially) weak. Other GMC properties like metallicity have weak effects (unless in some models with strong radiative feedback). Also, like the condition without BH feedback \citepalias{ShiKremerGrudic_2023MNRAS.518.3606S}, initial information about black holes like initial mass, position, and velocity does not significantly correlate with BH accretion (as long as BHs are gravitationally bound to the GMC initially), i.e., there is significant stochasticity in the process.
	
	\item With our fiducial feedback model (``log-form'' radiative efficiency and ``critical'' mechanical feedback efficiency, we find there is suppression of BH accretion in both simulations with $10^6\,\rm M_\odot$ and $10^8\,\rm M_\odot$ GMCs. Significant BH accretion is more likely in the $10^8\,\rm M_\odot$ GMC as expected, where BHs grow up to $\sim 5\times 10^4\,\rm M_\odot$, even though we assume relatively efficient radiative $\langle \epsilon_{\rm r} \rangle\sim 0.003$ and $\langle\epsilon_{\rm w} \rangle \sim 0.001$ (so $\langle \epsilon_{\rm w} \rangle /\langle \epsilon_{\rm r}\rangle \sim 1/3$). There are short phases of hyper-Eddington accretion ($f_{\rm Edd} \sim 1000$) for both simulations.
	
	\item For feedback physics involved in this study, we find that the maximum possible accreted mass for BHs $\Delta M_{\rm BH}^{\rm max}$
	declines as the feedback efficiency increases (Fig.~\ref{fig:feedback-scaling}), in a manner consistent with momentum-regulated BH growth. The simulations agree well with a simple analytic scaling summarized in Eq.~\eqref{equ:fitting-relation}. Depending on the cloud's properties and the feedback efficiency, the scenario can produce $10^4\,\rm M_\odot$ IMBHs quickly in a few free-fall times (typically $\sim \rm Myr$). These IMBHs can be the ``massive seeds'' of subsequent galactic SMBHs.
	
	\item For a given feedback efficiency model, the amount BHs can grow increases both with the surface density/acceleration scale of the parent complex, and the gas mass available. So dense, massive environments such as high-redshift galaxy nuclei will (unsurprisingly) be most favorable to rapid IMBH and even SMBH-level growth. Owing to the different physical mechanisms by which energy is converted into momentum, the same ``energetic'' efficiency $\epsilon$ for different physical mechanisms (radiation, mechanical outflows,  cosmic rays) will not produce the same level of self-regulation (Eq.~\ref{equ:fitting-relation}). Physically-motivated models of a variable radiative efficiency $\epsilon_{\rm r}$ \citep[e.g.][]{WataraiFukueTakeuchi_2000PASJ...52..133W,MadauHaardtDotti_2014ApJ...784L..38M} tend to predict relatively low $\epsilon_{r}$ at high Eddington ratios, so it seems unlikely that this is actually the dominant self-regulation mechanism on scales like what we resolve here. In contrast, some recent models for broad-opening angle, non-relativistic winds from super-Eddington disks \citep{HuInayoshiHaiman_2022ApJ...934..132H} have argued for this producing efficiently momentum-loaded winds with large ($\eta_{\rm w} \gg 1$) mass-loading, which would expel most of the gravitationally captured mass before it can reach the BH, and so in our simulations here much more efficiently suppress BH growth.
	
	\item Even when BH accretion overall is suppressed by BH feedback with high efficiency, there are still short phases of very high accretion rate ($\gtrsim 100\dot{M}_{\rm Edd}$) for a subset of BHs in the simulation ($\sim 10\% $ for the $10^6\,\rm M_\odot$ GMC and $\sim 1\%$ for the $10^8\,\rm M_\odot$ one). 
	
	\item When varying the BH feedback efficiency, we find that the massive complex's global star formation efficiency is generally not suppressed by BH feedback, unless the BH feedback efficiency is very high ($\gtrsim 0.01$) such that the could is disrupted within the first free-fall time. 
	
	\item Other parameters are related to the sub-grid (unresolved) mass flow to the BH, like the effective $\alpha$, the upper limit of BH accretion rate, and the jet mass loading factor $\eta_{\rm w}$ are largely degenerate, but do affect the BH mass growth (as expected by definition in our models).
\end{enumerate}

Some of these findings are similar to other recent works. For example, \citet{LupiHaardtDotti_2016MNRAS.456.2993L} simulated the super-critical accretion onto stellar-mass BH seeds in gaseous circum-nuclear disks (CNDs), with constant $\epsilon_{\rm r}=0.1$ or the slim disk model \citep[]{MadauHaardtDotti_2014ApJ...784L..38M} for radiative efficiency. \cite{PezzulliValianteSchneider_2016MNRAS.458.3047P} studied the growth of the central BH (initially with $\sim 100\,\rm M_\odot$ seed mass) of high redshift quasars with a semi-analytic framework including the slim-disk model and AGN wind feedback efficiency of $\sim 10^{-3}$ (but not CRs). Our results are similar to theirs for simulations with similar feedback parameters.

Extension of this work may include more sophisticated sub-grid models of BH accretion, which connect the BH accretion with the properties of the ambient gas. Another direction is to develop zoom-in simulations that attain higher resolution near the BH seed, which is more self-consistent but computationally more expensive. We leave these possibilities for future work.

\begin{acknowledgements}
      Support for the authors was provided
      by NSF Research Grants 1911233, 20009234, 2108318, NSF CAREER grant 1455342, NASA grants 80NSSC18K0562, HST-AR15800. Numerical calculations were run on the Caltech compute
      cluster “Wheeler,” allocations AST21010 and AST20016 supported
      by the NSF and TACC, and NASA HEC SMD-16-7592. A public version of the
      GIZMO code is available at \url{http://www.tapir.caltech.
      	edu/~phopkins/Site/GIZMO.html}. YS acknowledges the support of the Natural Sciences and Engineering Research Council of Canada (NSERC), [funding reference number 568580].
\end{acknowledgements}

%
\bibliographystyle{aa} 
\bibliography{new_bib} 

\begin{thebibliography}{83}
\expandafter\ifx\csname natexlab\endcsname\relax\def\natexlab#1{#1}\fi

\bibitem[{{Ba{\~n}ados} {et~al.}(2018){Ba{\~n}ados}, {Venemans},
  {Mazzucchelli}, {Farina}, {Walter}, {Wang}, {Decarli}, {Stern}, {Fan},
  {Davies}, \& et~al.}]{BanadosVenemansMazzucchelli_2018Natur.553..473B}
{Ba{\~n}ados}, E., {Venemans}, B.~P., {Mazzucchelli}, C., {et~al.} 2018, \nat,
  553, 473

\bibitem[{{Bate} {et~al.}(1995){Bate}, {Bonnell}, \&
  {Price}}]{BateBonnellPrice_1995MNRAS.277..362B}
{Bate}, M.~R., {Bonnell}, I.~A., \& {Price}, N.~M. 1995, \mnras, 277, 362

\bibitem[{{Begelman}(1979)}]{Begelman_1979MNRAS.187..237B}
{Begelman}, M.~C. 1979, \mnras, 187, 237

\bibitem[{{Blandford} \&
  {Begelman}(1999)}]{BlandfordBegelman_1999MNRAS.303L...1B}
{Blandford}, R.~D. \& {Begelman}, M.~C. 1999, \mnras, 303, L1

\bibitem[{{Blandford} \&
  {Begelman}(2004)}]{BlandfordBegelman_2004MNRAS.349...68B}
{Blandford}, R.~D. \& {Begelman}, M.~C. 2004, \mnras, 349, 68

\bibitem[{{Bondi}(1952)}]{Bondi_1952MNRAS.112..195B}
{Bondi}, H. 1952, \mnras, 112, 195

\bibitem[{{Bower} {et~al.}(2017){Bower}, {Schaye}, {Frenk}, {Theuns},
  {Schaller}, {Crain}, \& {McAlpine}}]{BowerSchayeFrenk_2017MNRAS.465...32B}
{Bower}, R.~G., {Schaye}, J., {Frenk}, C.~S., {et~al.} 2017, \mnras, 465, 32

\bibitem[{{Bromm} \& {Loeb}(2003)}]{BrommLoeb_2003ApJ...596...34B}
{Bromm}, V. \& {Loeb}, A. 2003, \apj, 596, 34

\bibitem[{{Chevance} {et~al.}(2023){Chevance}, {Krumholz}, {McLeod},
  {Ostriker}, {Rosolowsky}, \&
  {Sternberg}}]{ChevanceKrumholzMcLeod_2023ASPC..534....1C}
{Chevance}, M., {Krumholz}, M.~R., {McLeod}, A.~F., {et~al.} 2023, in
  Astronomical Society of the Pacific Conference Series, Vol. 534, Protostars
  and Planets VII, ed. S.~{Inutsuka}, Y.~{Aikawa}, T.~{Muto}, K.~{Tomida}, \&
  M.~{Tamura}, 1

\bibitem[{{Di Matteo} {et~al.}(2005){Di Matteo}, {Springel}, \&
  {Hernquist}}]{DiMatteoSpringelHernquist_2005Natur.433..604D}
{Di Matteo}, T., {Springel}, V., \& {Hernquist}, L. 2005, \nat, 433, 604

\bibitem[{{Dubois} {et~al.}(2013){Dubois}, {Pichon}, {Devriendt}, {Silk},
  {Haehnelt}, {Kimm}, \& {Slyz}}]{DuboisPichonDevriendt_2013MNRAS.428.2885D}
{Dubois}, Y., {Pichon}, C., {Devriendt}, J., {et~al.} 2013, \mnras, 428, 2885

\bibitem[{{Dubois} {et~al.}(2015){Dubois}, {Volonteri}, {Silk}, {Devriendt},
  {Slyz}, \& {Teyssier}}]{DuboisVolonteriSilk_2015MNRAS.452.1502D}
{Dubois}, Y., {Volonteri}, M., {Silk}, J., {et~al.} 2015, \mnras, 452, 1502

\bibitem[{{Fabian}(2012)}]{Fabian_2012ARA&A..50..455F}
{Fabian}, A.~C. 2012, \araa, 50, 455

\bibitem[{{Fukushima} {et~al.}(2020){Fukushima}, {Yajima}, {Sugimura},
  {Hosokawa}, {Omukai}, \&
  {Matsumoto}}]{FukushimaYajimaSugimura_2020MNRAS.497.3830F}
{Fukushima}, H., {Yajima}, H., {Sugimura}, K., {et~al.} 2020, \mnras, 497, 3830

\bibitem[{{Grudi{\'c}} {et~al.}(2018){Grudi{\'c}}, {Guszejnov}, {Hopkins},
  {Lamberts}, {Boylan-Kolchin}, {Murray}, \&
  {Schmitz}}]{GrudicGuszejnovHopkins_2018MNRAS.481..688G}
{Grudi{\'c}}, M.~Y., {Guszejnov}, D., {Hopkins}, P.~F., {et~al.} 2018, \mnras,
  481, 688

\bibitem[{{Grudi{\'c}} {et~al.}(2021{\natexlab{a}}){Grudi{\'c}}, {Guszejnov},
  {Hopkins}, {Offner}, \&
  {Faucher-Gigu{\`e}re}}]{GrudicGuszejnovHopkins_2021MNRAS.506.2199G}
{Grudi{\'c}}, M.~Y., {Guszejnov}, D., {Hopkins}, P.~F., {Offner}, S. S.~R., \&
  {Faucher-Gigu{\`e}re}, C.-A. 2021{\natexlab{a}}, \mnras, 506, 2199

\bibitem[{{Grudi{\'c}} {et~al.}(2021{\natexlab{b}}){Grudi{\'c}}, {Kruijssen},
  {Faucher-Gigu{\`e}re}, {Hopkins}, {Ma}, {Quataert}, \&
  {Boylan-Kolchin}}]{GrudicKruijssenFaucher-Giguere_2021MNRAS.506.3239G}
{Grudi{\'c}}, M.~Y., {Kruijssen}, J.~M.~D., {Faucher-Gigu{\`e}re}, C.-A.,
  {et~al.} 2021{\natexlab{b}}, \mnras, 506, 3239

\bibitem[{{Guo} \& {Mathews}(2012)}]{GuoMathews_2012ApJ...756..181G}
{Guo}, F. \& {Mathews}, W.~G. 2012, \apj, 756, 181

\bibitem[{{Guo} \& {Oh}(2008)}]{GuoOh_2008MNRAS.384..251G}
{Guo}, F. \& {Oh}, S.~P. 2008, \mnras, 384, 251

\bibitem[{{Guo} {et~al.}(2023){Guo}, {Stone}, {Kim}, \&
  {Quataert}}]{GuoStoneKim_2023ApJ...946...26G}
{Guo}, M., {Stone}, J.~M., {Kim}, C.-G., \& {Quataert}, E. 2023, \apj, 946, 26

\bibitem[{{Guszejnov} {et~al.}(2021){Guszejnov}, {Grudi{\'c}}, {Hopkins},
  {Offner}, \&
  {Faucher-Gigu{\`e}re}}]{GuszejnovGrudicHopkins_2021MNRAS.502.3646G}
{Guszejnov}, D., {Grudi{\'c}}, M.~Y., {Hopkins}, P.~F., {Offner}, S. S.~R., \&
  {Faucher-Gigu{\`e}re}, C.-A. 2021, \mnras, 502, 3646

\bibitem[{{Habouzit} {et~al.}(2017){Habouzit}, {Volonteri}, \&
  {Dubois}}]{HabouzitVolonteriDubois_2017MNRAS.468.3935H}
{Habouzit}, M., {Volonteri}, M., \& {Dubois}, Y. 2017, \mnras, 468, 3935

\bibitem[{{He} {et~al.}(2019){He}, {Ricotti}, \&
  {Geen}}]{HeRicottiGeen_2019MNRAS.489.1880H}
{He}, C.-C., {Ricotti}, M., \& {Geen}, S. 2019, \mnras, 489, 1880

\bibitem[{{Hopkins}(2015)}]{Hopkins_2015MNRAS.450...53H}
{Hopkins}, P.~F. 2015, \mnras, 450, 53

\bibitem[{{Hopkins}(2016)}]{Hopkins_2016MNRAS.462..576H}
{Hopkins}, P.~F. 2016, \mnras, 462, 576

\bibitem[{{Hopkins} {et~al.}(2021){Hopkins}, {Chan}, {Ji}, {Hummels},
  {Kere{\v{s}}}, {Quataert}, \&
  {Faucher-Gigu{\`e}re}}]{HopkinsChanJi_2021MNRAS.501.3640H}
{Hopkins}, P.~F., {Chan}, T.~K., {Ji}, S., {et~al.} 2021, \mnras, 501, 3640

\bibitem[{{Hopkins} {et~al.}(2020){Hopkins}, {Grudi{\'c}}, {Wetzel},
  {Kere{\v{s}}}, {Faucher-Gigu{\`e}re}, {Ma}, {Murray}, \&
  {Butcher}}]{HopkinsGrudicWetzel_2020MNRAS.491.3702H}
{Hopkins}, P.~F., {Grudi{\'c}}, M.~Y., {Wetzel}, A., {et~al.} 2020, \mnras,
  491, 3702

\bibitem[{{Hopkins} \& {Raives}(2016)}]{HopkinsRaives_2016MNRAS.455...51H}
{Hopkins}, P.~F. \& {Raives}, M.~J. 2016, \mnras, 455, 51

\bibitem[{{Hopkins} {et~al.}(2022){Hopkins}, {Squire}, \&
  {Butsky}}]{HopkinsSquireButsky_2022MNRAS.509.3779H}
{Hopkins}, P.~F., {Squire}, J., \& {Butsky}, I.~S. 2022, \mnras, 509, 3779

\bibitem[{{Hopkins} {et~al.}(2024{\natexlab{a}}){Hopkins}, {Squire},
  {Quataert}, {Murray}, {Su}, {Steinwandel}, {Kremer}, {Faucher-Giguere}, \&
  {Wellons}}]{HopkinsSquireQuataert_2024OJAp....7E..20H}
{Hopkins}, P.~F., {Squire}, J., {Quataert}, E., {et~al.} 2024{\natexlab{a}},
  The Open Journal of Astrophysics, 7, 20

\bibitem[{{Hopkins} {et~al.}(2024{\natexlab{b}}){Hopkins}, {Squire}, {Su},
  {Steinwandel}, {Kremer}, {Shi}, {Grudic}, {Wellons}, {Faucher-Giguere},
  {Angles-Alcazar}, \& et~al.}]{HopkinsSquireSu_2024OJAp....7E..19H}
{Hopkins}, P.~F., {Squire}, J., {Su}, K.-Y., {et~al.} 2024{\natexlab{b}}, The
  Open Journal of Astrophysics, 7, 19

\bibitem[{{Hopkins} {et~al.}(2016){Hopkins}, {Torrey}, {Faucher-Gigu{\`e}re},
  {Quataert}, \& {Murray}}]{HopkinsTorreyFaucher-Giguere_2016MNRAS.458..816H}
{Hopkins}, P.~F., {Torrey}, P., {Faucher-Gigu{\`e}re}, C.-A., {Quataert}, E.,
  \& {Murray}, N. 2016, \mnras, 458, 816

\bibitem[{{Hopkins} {et~al.}(2018{\natexlab{a}}){Hopkins}, {Wetzel},
  {Kere{\v{s}}}, {Faucher-Gigu{\`e}re}, {Quataert}, {Boylan-Kolchin}, {Murray},
  {Hayward}, \& {El-Badry}}]{HopkinsWetzelKeres_2018MNRAS.477.1578H}
{Hopkins}, P.~F., {Wetzel}, A., {Kere{\v{s}}}, D., {et~al.} 2018{\natexlab{a}},
  \mnras, 477, 1578

\bibitem[{{Hopkins} {et~al.}(2018{\natexlab{b}}){Hopkins}, {Wetzel},
  {Kere{\v{s}}}, {Faucher-Gigu{\`e}re}, {Quataert}, {Boylan-Kolchin}, {Murray},
  {Hayward}, {Garrison-Kimmel}, {Hummels}, \&
  et~al.}]{HopkinsWetzelKeres_2018MNRAS.480..800H}
{Hopkins}, P.~F., {Wetzel}, A., {Kere{\v{s}}}, D., {et~al.} 2018{\natexlab{b}},
  \mnras, 480, 800

\bibitem[{{Hopkins} {et~al.}(2023){Hopkins}, {Wetzel}, {Wheeler}, {Sanderson},
  {Grudi{\'c}}, {Sameie}, {Boylan-Kolchin}, {Orr}, {Ma}, {Faucher-Gigu{\`e}re},
  \& et~al.}]{HopkinsWetzelWheeler_2023MNRAS.519.3154H}
{Hopkins}, P.~F., {Wetzel}, A., {Wheeler}, C., {et~al.} 2023, \mnras, 519, 3154

\bibitem[{{Hoyle} \& {Lyttleton}(1939)}]{HoyleLyttleton_1939PCPS...35..405H}
{Hoyle}, F. \& {Lyttleton}, R.~A. 1939, Proceedings of the Cambridge
  Philosophical Society, 35, 405

\bibitem[{{Hu} {et~al.}(2022){Hu}, {Inayoshi}, {Haiman}, {Quataert}, \&
  {Kuiper}}]{HuInayoshiHaiman_2022ApJ...934..132H}
{Hu}, H., {Inayoshi}, K., {Haiman}, Z., {Quataert}, E., \& {Kuiper}, R. 2022,
  \apj, 934, 132

\bibitem[{{Inayoshi} {et~al.}(2016){Inayoshi}, {Haiman}, \&
  {Ostriker}}]{InayoshiHaimanOstriker_2016MNRAS.459.3738I}
{Inayoshi}, K., {Haiman}, Z., \& {Ostriker}, J.~P. 2016, \mnras, 459, 3738

\bibitem[{{Inayoshi} {et~al.}(2020){Inayoshi}, {Visbal}, \&
  {Haiman}}]{InayoshiVisbalHaiman_2020ARA&A..58...27I}
{Inayoshi}, K., {Visbal}, E., \& {Haiman}, Z. 2020, \araa, 58, 27

\bibitem[{{Ishibashi} \& {Fabian}(2023)}]{IshibashiFabian_2023MNRAS.519.1931I}
{Ishibashi}, W. \& {Fabian}, A.~C. 2023, \mnras, 519, 1931

\bibitem[{{Ji} {et~al.}(2020){Ji}, {Chan}, {Hummels}, {Hopkins}, {Stern},
  {Kere{\v{s}}}, {Quataert}, {Faucher-Gigu{\`e}re}, \&
  {Murray}}]{JiChanHummels_2020MNRAS.496.4221J}
{Ji}, S., {Chan}, T.~K., {Hummels}, C.~B., {et~al.} 2020, \mnras, 496, 4221

\bibitem[{{Jiang} {et~al.}(2014){Jiang}, {Stone}, \&
  {Davis}}]{JiangStoneDavis_2014ApJ...796..106J}
{Jiang}, Y.-F., {Stone}, J.~M., \& {Davis}, S.~W. 2014, \apj, 796, 106

\bibitem[{{Jiang} {et~al.}(2019){Jiang}, {Stone}, \&
  {Davis}}]{JiangStoneDavis_2019ApJ...880...67J}
{Jiang}, Y.-F., {Stone}, J.~M., \& {Davis}, S.~W. 2019, \apj, 880, 67

\bibitem[{{Kim} {et~al.}(2018){Kim}, {Kim}, \&
  {Ostriker}}]{KimKimOstriker_2018ApJ...859...68K}
{Kim}, J.-G., {Kim}, W.-T., \& {Ostriker}, E.~C. 2018, \apj, 859, 68

\bibitem[{{Kim} {et~al.}(2021){Kim}, {Ostriker}, \&
  {Filippova}}]{KimOstrikerFilippova_2021ApJ...911..128K}
{Kim}, J.-G., {Ostriker}, E.~C., \& {Filippova}, N. 2021, \apj, 911, 128

\bibitem[{{Klessen}(2000)}]{Klessen_2000ApJ...535..869K}
{Klessen}, R.~S. 2000, \apj, 535, 869

\bibitem[{{Kremer} {et~al.}(2020){Kremer}, {Spera}, {Becker}, {Chatterjee}, {Di
  Carlo}, {Fragione}, {Rodriguez}, {Ye}, \&
  {Rasio}}]{KremerSperaBecker_2020ApJ...903...45K}
{Kremer}, K., {Spera}, M., {Becker}, D., {et~al.} 2020, \apj, 903, 45

\bibitem[{{Krumholz} {et~al.}(2020){Krumholz}, {Crocker}, {Xu}, {Lazarian},
  {Rosevear}, \& {Bedwell-Wilson}}]{KrumholzCrockerXu_2020MNRAS.493.2817K}
{Krumholz}, M.~R., {Crocker}, R.~M., {Xu}, S., {et~al.} 2020, \mnras, 493, 2817

\bibitem[{{Larson} \&
  {Starrfield}(1971)}]{LarsonStarrfield_1971A&A....13..190L}
{Larson}, R.~B. \& {Starrfield}, S. 1971, \aap, 13, 190

\bibitem[{{Latif} {et~al.}(2022){Latif}, {Whalen}, {Khochfar}, {Herrington}, \&
  {Woods}}]{LatifWhalenKhochfar_2022Natur.607...48L}
{Latif}, M.~A., {Whalen}, D.~J., {Khochfar}, S., {Herrington}, N.~P., \&
  {Woods}, T.~E. 2022, \nat, 607, 48

\bibitem[{{Lupi} {et~al.}(2016){Lupi}, {Haardt}, {Dotti}, {Fiacconi}, {Mayer},
  \& {Madau}}]{LupiHaardtDotti_2016MNRAS.456.2993L}
{Lupi}, A., {Haardt}, F., {Dotti}, M., {et~al.} 2016, \mnras, 456, 2993

\bibitem[{{Lupi} {et~al.}(2024){Lupi}, {Quadri}, {Volonteri}, {Colpi}, \&
  {Regan}}]{LupiQuadriVolonteri_2024A&A...686A.256L}
{Lupi}, A., {Quadri}, G., {Volonteri}, M., {Colpi}, M., \& {Regan}, J.~A. 2024,
  \aap, 686, A256

\bibitem[{{Mac Low} \& {Klessen}(2004)}]{MacLowKlessen_2004RvMP...76..125M}
{Mac Low}, M.-M. \& {Klessen}, R.~S. 2004, Reviews of Modern Physics, 76, 125

\bibitem[{{Madau} {et~al.}(2014){Madau}, {Haardt}, \&
  {Dotti}}]{MadauHaardtDotti_2014ApJ...784L..38M}
{Madau}, P., {Haardt}, F., \& {Dotti}, M. 2014, \apjl, 784, L38

\bibitem[{{Madau} \& {Rees}(2001)}]{MadauRees_2001ApJ...551L..27M}
{Madau}, P. \& {Rees}, M.~J. 2001, \apjl, 551, L27

\bibitem[{{Massonneau} {et~al.}(2023){Massonneau}, {Volonteri}, {Dubois}, \&
  {Beckmann}}]{MassonneauVolonteriDubois_2023A&A...670A.180M}
{Massonneau}, W., {Volonteri}, M., {Dubois}, Y., \& {Beckmann}, R.~S. 2023,
  \aap, 670, A180

\bibitem[{{McKee} \& {Ostriker}(2007)}]{McKeeOstriker_2007ARA&A..45..565M}
{McKee}, C.~F. \& {Ostriker}, E.~C. 2007, \araa, 45, 565

\bibitem[{{Murray} {et~al.}(2005){Murray}, {Quataert}, \&
  {Thompson}}]{MurrayQuataertThompson_2005ApJ...618..569M}
{Murray}, N., {Quataert}, E., \& {Thompson}, T.~A. 2005, \apj, 618, 569

\bibitem[{{Ohsuga} {et~al.}(2005){Ohsuga}, {Mori}, {Nakamoto}, \&
  {Mineshige}}]{OhsugaMoriNakamoto_2005ApJ...628..368O}
{Ohsuga}, K., {Mori}, M., {Nakamoto}, T., \& {Mineshige}, S. 2005, \apj, 628,
  368

\bibitem[{{Ostriker} {et~al.}(2010){Ostriker}, {Choi}, {Ciotti}, {Novak}, \&
  {Proga}}]{OstrikerChoiCiotti_2010ApJ...722..642O}
{Ostriker}, J.~P., {Choi}, E., {Ciotti}, L., {Novak}, G.~S., \& {Proga}, D.
  2010, \apj, 722, 642

\bibitem[{{Pezzulli} {et~al.}(2016){Pezzulli}, {Valiante}, \&
  {Schneider}}]{PezzulliValianteSchneider_2016MNRAS.458.3047P}
{Pezzulli}, E., {Valiante}, R., \& {Schneider}, R. 2016, \mnras, 458, 3047

\bibitem[{{Portegies Zwart} {et~al.}(2004){Portegies Zwart}, {Baumgardt},
  {Hut}, {Makino}, \&
  {McMillan}}]{PortegiesZwartBaumgardtHut_2004Natur.428..724P}
{Portegies Zwart}, S.~F., {Baumgardt}, H., {Hut}, P., {Makino}, J., \&
  {McMillan}, S. L.~W. 2004, \nat, 428, 724

\bibitem[{{Quataert} \&
  {Gruzinov}(2000)}]{QuataertGruzinov_2000ApJ...539..809Q}
{Quataert}, E. \& {Gruzinov}, A. 2000, \apj, 539, 809

\bibitem[{{Regan} {et~al.}(2019){Regan}, {Downes}, {Volonteri}, {Beckmann},
  {Lupi}, {Trebitsch}, \& {Dubois}}]{ReganDownesVolonteri_2019MNRAS.486.3892R}
{Regan}, J.~A., {Downes}, T.~P., {Volonteri}, M., {et~al.} 2019, \mnras, 486,
  3892

\bibitem[{{Ryu} {et~al.}(2016){Ryu}, {Tanaka}, {Perna}, \&
  {Haiman}}]{RyuTanakaPerna_2016MNRAS.460.4122R}
{Ryu}, T., {Tanaka}, T.~L., {Perna}, R., \& {Haiman}, Z. 2016, \mnras, 460,
  4122

\bibitem[{{Shakura} \& {Sunyaev}(1973)}]{ShakuraSunyaev_1973A&A....24..337S}
{Shakura}, N.~I. \& {Sunyaev}, R.~A. 1973, \aap, 24, 337

\bibitem[{{Shen} {et~al.}(2020){Shen}, {Hopkins}, {Faucher-Gigu{\`e}re},
  {Alexander}, {Richards}, {Ross}, \&
  {Hickox}}]{ShenHopkinsFaucher-Giguere_2020MNRAS.495.3252S}
{Shen}, X., {Hopkins}, P.~F., {Faucher-Gigu{\`e}re}, C.-A., {et~al.} 2020,
  \mnras, 495, 3252

\bibitem[{{Shi} {et~al.}(2021){Shi}, {Grudi{\'c}}, \&
  {Hopkins}}]{ShiGrudicHopkins_2021MNRAS.505.2753S}
{Shi}, Y., {Grudi{\'c}}, M.~Y., \& {Hopkins}, P.~F. 2021, \mnras, 505, 2753

\bibitem[{{Shi} {et~al.}(2023){Shi}, {Kremer}, {Grudi{\'c}},
  {Gerling-Dunsmore}, \& {Hopkins}}]{ShiKremerGrudic_2023MNRAS.518.3606S}
{Shi}, Y., {Kremer}, K., {Grudi{\'c}}, M.~Y., {Gerling-Dunsmore}, H.~J., \&
  {Hopkins}, P.~F. 2023, \mnras, 518, 3606

\bibitem[{{Sijacki} {et~al.}(2008){Sijacki}, {Pfrommer}, {Springel}, \&
  {En{\ss}lin}}]{SijackiPfrommerSpringel_2008MNRAS.387.1403S}
{Sijacki}, D., {Pfrommer}, C., {Springel}, V., \& {En{\ss}lin}, T.~A. 2008,
  \mnras, 387, 1403

\bibitem[{{Silk} \& {Rees}(1998)}]{SilkRees_1998A&A...331L...1S}
{Silk}, J. \& {Rees}, M.~J. 1998, \aap, 331, L1

\bibitem[{{S{\k{a}}dowski}(2009)}]{Sadowski_2009ApJS..183..171S}
{S{\k{a}}dowski}, A. 2009, \apjs, 183, 171

\bibitem[{{S{\k{a}}dowski} {et~al.}(2016){S{\k{a}}dowski}, {Lasota},
  {Abramowicz}, \& {Narayan}}]{SadowskiLasotaAbramowicz_2016MNRAS.456.3915S}
{S{\k{a}}dowski}, A., {Lasota}, J.-P., {Abramowicz}, M.~A., \& {Narayan}, R.
  2016, \mnras, 456, 3915

\bibitem[{{S{\k{a}}dowski} {et~al.}(2015){S{\k{a}}dowski}, {Narayan},
  {Tchekhovskoy}, {Abarca}, {Zhu}, \&
  {McKinney}}]{SadowskiNarayanTchekhovskoy_2015MNRAS.447...49S}
{S{\k{a}}dowski}, A., {Narayan}, R., {Tchekhovskoy}, A., {et~al.} 2015, \mnras,
  447, 49

\bibitem[{{Su} {et~al.}(2021){Su}, {Hopkins}, {Bryan}, {Somerville}, {Hayward},
  {Angl{\'e}s-Alc{\'a}zar}, {Faucher-Gigu{\`e}re}, {Wellons}, {Stern},
  {Terrazas}, \& et~al.}]{SuHopkinsBryan_2021MNRAS.507..175S}
{Su}, K.-Y., {Hopkins}, P.~F., {Bryan}, G.~L., {et~al.} 2021, \mnras, 507, 175

\bibitem[{{Talbot} {et~al.}(2021){Talbot}, {Bourne}, \&
  {Sijacki}}]{TalbotBourneSijacki_2021MNRAS.504.3619T}
{Talbot}, R.~Y., {Bourne}, M.~A., \& {Sijacki}, D. 2021, \mnras, 504, 3619

\bibitem[{{Torrey} {et~al.}(2020){Torrey}, {Hopkins}, {Faucher-Gigu{\`e}re},
  {Angl{\'e}s-Alc{\'a}zar}, {Quataert}, {Ma}, {Feldmann}, {Keres}, \&
  {Murray}}]{TorreyHopkinsFaucher-Giguere_2020MNRAS.497.5292T}
{Torrey}, P., {Hopkins}, P.~F., {Faucher-Gigu{\`e}re}, C.-A., {et~al.} 2020,
  \mnras, 497, 5292

\bibitem[{{Volonteri} {et~al.}(2021){Volonteri}, {Habouzit}, \&
  {Colpi}}]{VolonteriHabouzitColpi_2021NatRP...3..732V}
{Volonteri}, M., {Habouzit}, M., \& {Colpi}, M. 2021, Nature Reviews Physics,
  3, 732

\bibitem[{{Wang} {et~al.}(2021){Wang}, {Yang}, {Fan}, {Hennawi}, {Barth},
  {Banados}, {Bian}, {Boutsia}, {Connor}, {Davies}, \&
  et~al.}]{WangYangFan_2021ApJ...907L...1W}
{Wang}, F., {Yang}, J., {Fan}, X., {et~al.} 2021, \apjl, 907, L1

\bibitem[{{Watarai} {et~al.}(2000){Watarai}, {Fukue}, {Takeuchi}, \&
  {Mineshige}}]{WataraiFukueTakeuchi_2000PASJ...52..133W}
{Watarai}, K.-y., {Fukue}, J., {Takeuchi}, M., \& {Mineshige}, S. 2000, \pasj,
  52, 133

\bibitem[{{Yang} {et~al.}(2020){Yang}, {Wang}, {Fan}, {Hennawi}, {Davies},
  {Yue}, {Banados}, {Wu}, {Venemans}, {Barth}, \&
  et~al.}]{YangWangFan_2020ApJ...897L..14Y}
{Yang}, J., {Wang}, F., {Fan}, X., {et~al.} 2020, \apjl, 897, L14

\bibitem[{{Yu} \& {Tremaine}(2002)}]{YuTremaine_2002MNRAS.335..965Y}
{Yu}, Q. \& {Tremaine}, S. 2002, \mnras, 335, 965

\bibitem[{{Zweibel}(2017)}]{Zweibel_2017PhPl...24e5402Z}
{Zweibel}, E.~G. 2017, Physics of Plasmas, 24, 055402

\end{thebibliography}
%

\end{document}